\documentclass{article}

\usepackage{amssymb,amsfonts,amsmath,amsthm,lineno,mathrsfs,mathtools}

\MakeRobust{\eqref}

\usepackage{hyperref}
\usepackage{enumitem}

\newcommand{\mR}{\mathbb{R}}
\newcommand{\mZ}{\mathbb{Z}}

\newcommand{\lb}{\mathbf{l}}
\newcommand{\vv}{\mathbf{v}}

\newcommand{\dsp}{\displaystyle}
\newcommand{\txt}{\textstyle}
\newcommand{\al}{\alpha}

\newcommand{\vep}{\varepsilon}
\newcommand{\la}{\lambda}
\newcommand{\La}{\Lambda}
\newcommand{\p}{\partial}
\newcommand{\w}{\omega}
\newcommand{\W}{\Omega}
\newcommand{\Ac}{\mathcal{A}}

\newcommand{\Bc}{\mathcal{B}}

\newcommand{\Dc}{\mathcal{D}}

\newcommand{\Hc}{\mathcal{H}}

\newcommand{\Oc}{\mathcal{O}}
\newcommand{\ov}{\overline}
\newcommand{\con}{\mathrm{const}}
\newcommand{\1}{\mathbf{1}}

\newcommand{\inc}{\mathrm{in}}
\newcommand{\out}{\mathrm{out}}
\newcommand{\ret}{\mathrm{ret}}
\newcommand{\adv}{\mathrm{adv}}
\newcommand{\rad}{\mathrm{rad}}
\newcommand{\fre}{\mathrm{f}}
\newcommand{\as}{\mathrm{as}}

\newcommand{\qu}{\mathrm{q}}
\newcommand{\dW}{\dot{W}}
\newcommand{\dV}{\dot{V}}
\newcommand{\ddV}{\ddot{V}}

\newcommand{\zb}{\bar{z}}
\newcommand{\ti}{\widetilde}

\DeclareMathOperator{\Ip}{Im}
\DeclareMathOperator{\id}{id}

\DeclareMathOperator{\sgn}{sgn}

\usepackage[a4paper, total={15.6cm, 25cm}]{geometry}

\title{Infrared structure beyond locality in electrodynamics}
\author{Andrzej Herdegen\thanks{e-mail: herdegen@th.if.uj.edu.pl}\\
{\it Institute of Theoretical Physics, Jagiellonian University,}\\
{\it ul.\,S.\,{\L}ojasiewicza 11, 30-348  Krak\'{o}w, Poland}}\date{}

\begin{document}

\maketitle

\begin{abstract}
The infrared problems of quantum electrodynamics, in contrast to ultraviolet difficulties which are of technical nature, are related to fundamental, conceptual physical questions, such as: what is a charged particle, is the particle interpretation of the electromagnetic field complete, does a vacuum state exist, or what is the quantum status of long range degrees of freedom. On the calculational level, the standard local formulations of quantum field theory have achieved procedures to deal with infinities related to long range correlations. However, the answers to the conceptual questions formulated above, based on the locality paradigm, do not seem to be fully convincing, which is confirmed by the fact that no canonical picture did emerge. This contribution briefly characterizes perspectives which open with an admission of nonlocal variables residing in infinity, or at the boundary of spacetime after compactification. Recently, this line of investigation gains popularity.

\end{abstract}

\section{Introduction}

The standard formulations of quantum field theory include, as one of their fundamental paradigms, the axiom of locality. This axiom may be  characterized by the following two statements.
\begin{itemize}[topsep=1ex,itemsep=0ex,leftmargin=2em]
\item[(1)] Basic observables of QFT models are labelled by bounded spacetime regions in which they may be measured.
\item[(2)] If observables $A_1$ and $A_2$ are labelled in this way by $\Oc_1$ and $\Oc_2$, respectively, and these regions are in spacelike position to each other, then $[A_1,A_2]=0$.
\end{itemize}
Both in the Wightman axiomatic formulation, and in the perturbative calculations, the dependence indicated in (1) takes the form of an operator valued distribution $A(\chi)$, where $\chi(x)$ is a test function with a compact support in the corresponding region $\Oc$. In the algebraic formulation this dependence takes a more abstract form of a net of local algebras.

From the fundamental, algebraic point of view, the physical content of a theory is fully contained in the algebra of observables, and a particular physical setting described by the theory corresponds to a choice of a representation of this algebra. A class of unitarily equivalent irreducible representations forms what is called a superselection sector of the theory.
However, for the actual construction of quantum field theory models, it proved extremely helpful, not to say indispensable, to consider not only observables, but also fields---operator valued distributions---which are not directly measurable, such as the Dirac field of electrons/positrons.\footnote{This has been recognized, in recent decades, by the algebraic approach to QFT as well, see a recent summary \cite{rej16}.}  For reasons which seem mainly of technical nature---manageability of the perturbative expansion and treatment of the ultraviolet problems---the locality axiom has been extended to include such fields as well, with the modification that in case of two fermion fields the commutator in (2) is replaced by the anticommutator.

The locality paradigm, fruitful as it is, faces problems in its actual constructive implementation in theories with constraints and long-range interaction, such as the emblematic example of quantum electrodynamics, which concerns us here. At the heart of these problems is the existence of long range correlations. The most immediate consequence of the Gauss law is that in a physically realistic representation of quantum electrodynamics charged fields cannot be local \cite{fer74}. This is easy to understand: if Maxwell's equations are satisfied, and the charge operator may be constructed by some regularized integration of the electric field on a large sphere, then with the radius of that sphere tending to infinity it commutes with all local fields; conclusion: such fields cannot create charge. More than that, it turns out that for the implementation of locality one has to admit indefinite metric spaces (Krein spaces) \cite{fer74}. Thus if one insists on locality, one has to agree that the construction may only be a first technical step on the road to a realistic theory. In some approaches the problem is solved by ignoring the question of spacelike infinity; the algebraic adiabatic limit technique for the construction of local algebras, or the approach restricting observable possibilities to future light cones, are in this class (see discussion in \cite{duc23}). However, here we are interested in treating long range correlations realistically, so we go back once more to a general analysis.

Consider a scattering situation, in which both the incoming and the outgoing electromagnetic currents are due to free massive particles. In this context it is useful to decompose the total electromagnetic field in two alternative ways, $F=F^\ret+F^\inc=F^\adv+F^\out$, where $F^\ret/F^\adv$ is the retarded/advanced field of the current, and $F^\inc/F^\out$ is a free incoming/outgoing field. These decompositions may be regarded as definitions of the respective free parts, but they also have a clear physical interpretation. Due to the propagation of $F^\ret$ from the source current along future light cones, this field approaches, both in far past, as well as in spacelike infinity,  the Coulomb field of the incoming particles; this justifies the interpretation of $F^\inc$ as a free incoming radiation field. Similar remarks and interpretation apply to $F^\adv/F^\out$, with interchange of future to past light cones.

Coulomb fields of free charges have spacelike tails of decay order $r^{-2}$. Therefore, unless the scattering is trivial, the spacelike tail of $F^\out-F^\inc=F^\ret-F^\adv$ is of the same order. It is thus consistent to assume that in general the decay of $F^\inc$, and then also of $F^\out$, is of the same order. Free fields with such spacelike decay will be called `infrared singular', while those decaying faster -- infrared regular: justification for that will become obvious later. In consequence, the total field decays at the same rate $r^{-2}$. More precisely, for each spacetime position vector $x$ and each spacelike vector $y$, one should have\footnote{Spacetime indices (which will frequently be suppressed) are denoted by $a,b,c$, etc., and the spacetime metric signature is $(+,-,-,-)$. We assume that a reference point in spacetime has been chosen, and then points are represented by vectors $x$, $y$, etc.\ in Minkowski vector space $M$. Also, we use units with $\hbar=1$, $c=1$.}
\begin{equation}\label{int_space}
 \lim_{r\to\infty}r^2F_{ab}(x+ry)=\lim_{r\to\infty}r^2F_{ab}^\ret(x+ry)
 +\lim_{r\to\infty}r^2F_{ab}^\inc(x+ry)=F_{ab}^0(y)\,,\qquad y^2<0\,,
\end{equation}
where the field $F_{ab}^0(y)$ is homogeneous of degree $-2$, independent of $x$ (more on this last property below). However, as $x+ry$ becomes, for large $r$, spacelike to any compact region in spacetime, in the local quantum electrodynamics $F^0(y)$ commutes with all basic observables, so it must be a numerical function, characterizing a given superselection sector. But then, it is hard to avoid the conclusion that the tail of the incoming field $F^\inc$ plays only a slave role with respect to the Coulomb fields of the particles, compensating their tails to the value prescribed by $F^0(y)$. Various ways to deal with this question have been proposed, most notably with the use of coherent representations of the free electromagnetic fields, and the dressing in the style proposed by Faddeev and Kulish, which involves introduction of a classical external current.\footnote{The most recent and complete implementation of the Faddeev-Kulish idea has been achieved in \cite{duc19} with the use of truncation of the interaction lagrangian and subsequent adiabatic limit, see also \cite{duc23}. This resulted in the construction of the infrared-finite low orders of the scattering operator, with conjectured extension to all orders. However, the charged fields do not exist as operators in the adiabatic limit, so the representation space of the truncated interaction is probably not that of the limit theory.}  However, in all of them, the free fields have to be correlated in this way, fundamentally hard to explain, to the particle degrees of freedom. Again, similar remarks apply to the `out' perspective (the lhs of \eqref{int_space} and $F^0_{ab}(y)$ remain the same).

There are two mathematical consequences of the above superselection structure: the obvious Lorentz symmetry breaking (there are no Lorentz invariant classical fields $F^0_{ab}(y)$), and the lack of a discrete value in the mass spectrum, corresponding to a free massive charged particle \cite{buch86}. The second consequence, known as the infraparticle problem, may be understood in terms of the `dressing' mentioned above: the accompanying photons blur the mass hyperboloid. There are also further suggestions which seem to follow from the above picture:
\begin{itemize}[topsep=1ex,itemsep=0ex,leftmargin=1.5em]
\item[--] the form of the long range tail of the electromagnetic field $F^0(y)$ is to a large extent a question of convention, and
\item[--] the infrared degrees of freedom of a free field differ fundamentally from their local characteristics.
\end{itemize}
However, one can give arguments to the contrary.
\begin{itemize}[topsep=1ex,itemsep=0ex,leftmargin=1.5em]
\item[--] Scattering events in quantum electrodynamics have the spacetime scale of a laboratory, and distances of order of meters are sufficient to be regraded as infinity. Moreover, the memory effect, discovered in electrodynamics by Staruszkiewicz, reveals observable consequences of the long range tails.
\item[--] Classically, free fields taking part in scattering are necessarily, in nontrivial contexts, infrared singular,  and all their degrees of freedom are fully autonomous. In quantum theory, the algebra of free local fields has a natural extension including infrared singular fields.
\end{itemize}

The local, and more orthodox, views on the infrared problem and its possible solutions are discussed in another article in this Encyclopedia \cite{duc23}. In this contribution we review the efforts to give the long range variables in electrodynamics fully autonomous and quantum character. It seems that the first, and at the same time the most radical proposition of this type was formulated by Staruszkiewicz in his theory of the quantum Coulomb field. Later followed a thorough analysis of long range properties of classical electrodynamics, and quantum constructions taking into account the results of this structure. Recently, elements of this program have been rediscovered by other authors, and some new structures and relations have been proposed.

In the following, we devote relatively large space to the classical structure, Section \ref{free}. This gives the necessary background for a brief summary of quantum ideas in Sections \ref{quant} and \ref{star}. The classical structure stands on firm ground, while the quantum ideas inspired by it are of more speculative nature. Our discussion is rather sketchy, precise assumptions and details of proofs may be found in the bibliography.

\section{Classical electromagnetic fields}\label{free}

\subsection{Light cone geometry}\label{lightcone}

We denote $C_+=\{l\in M|\,l^2=0, l^0>0\}$, the future light cone. There is a natural representation of the proper Lorentz group $\mathcal{L}_+^\uparrow$ acting on the space of functions on $C_+$, for $\La\in\mathcal{L}_+^\uparrow$ defined by $[T_\La f](l)=f(\La^{-1}l)$. Its generator is easily found: for $\La^a{}_b=\exp(\w^a{}_b)$, $\w_{ab}=-\w_{ba}$, and smooth $f$, one has
\begin{equation}
 \frac{\delta}{\delta\w^{ab}}[T_\La f](l)\big|_{\w^{ab}=0}
 =\tfrac{1}{2}L_{ab}f(l)\,,\qquad
 L_{ab}=l_a\p_b-l_b\p_a\,,
\end{equation}
where $\p_a=\p/\p l^a$, and for the sake of differentiation in $L_{ab}f$ any smooth extension of $f$ to a neighborhood of $C_+$ may be used---the result on the cone does not depend on the choice, as the lhs shows. Therefore, $L_{ab}$ is an intrinsic differentiation operator in the light cone.

In some cases we shall need a choice of a time axis in Minkowski space, and then we denote by $t$ the unit, future-pointing vector along this axis. Let $f(l)$ be a smooth function, homogeneous of degree $-2$:
$f(\la l)=\la^{-2}f(l)$, $\la>0$. Then the formula
\begin{equation}
 \int f(l)\,d^2l\coloneqq \int f(1,\lb)\,d\W(\lb)\,,
\end{equation}
where on the rhs $l$'s are scaled to $t\cdot l=1$ and $d\W(\lb)$ is the solid angle measure of the $3$-space (orthogonal to $t$) part $\lb$ of $l$, defines a Lorentz-invariant quantity (independent of the choice of $t$). This invariance implies, in particular,
\begin{equation}\label{Lf0}
 \int L_{ab}f(l)\,d^2l=0\,.
\end{equation}
If $V^a(l)$ is a vector function orthogonal to $l$, $l\cdot V(l)=0$, homogeneous of degree $-1$, and extended into a neighborhood of the cone with the preservation of these two properties, then for each $t$ one has $L_{ab}[t^aV^b(l)/t\cdot l]=\p\cdot V(l)$; in particular, this occurs for $V_a=\phi\p_a\psi(l)$ with $\phi(l)$, $\psi(l)$ homogeneous in a neighborhood of the cone. Therefore, the restrictions of $\p\cdot V(l)$ and of $\p\cdot [\phi(l)\p\psi(l)]$ to the cone do not depend on a choice of such extensions, and the following is satisfied
\begin{equation}\label{pV0}
 \int \p\cdot V(l) \,d^2l=0\,,\qquad
 \int \phi(l)\p^2\psi(l)\,d^2l=\int \psi(l)\p^2\phi(l)\,d^2l\,.
\end{equation}

Moreover, for each such $V(l)$ there exist scalar functions $\phi(l)$ and $\psi(l)$, homogeneous of degree $0$, such that
\begin{equation}\label{Vfp}
 l_aV_b(l)-l_bV_a(l)=L_{ab}\phi(l)-{}^*\!L_{ab}\psi(l)\,,
\end{equation}
where star ${}^*$ denotes the dual antisymmetric tensor (see Appendix of \cite{her98}). The scalar functions are determined up to the addition of a constant, and a special solution for $\phi$ is
\begin{equation}\label{fsp}
 \phi(l)=\frac{1}{4\pi}\int \frac{l\cdot V(l')}{l\cdot l'}\,d^2l'\,,
\end{equation}
which has the property that for each $t$ the following identity is satisfied
\begin{equation}\label{fspid}
 \int\frac{\phi(l)}{(t\cdot l)^2}\,d^2l=\int \frac{t\cdot V(l)}{t\cdot l}\,d^2l\,.
\end{equation}
Particular case of the relation \eqref{Vfp} is given by the following equivalence:
\begin{equation}\label{Vfi}
 l_aV_b(l)-l_bV_a(l)=L_{ab}\phi(l)\qquad \text{iff}\qquad L_{[ab}V_{c]}(l)=0\,.
\end{equation}

The cone $C_+$ is a three-parameter hypersurface. For any parametrization, with $\xi$ any of the parameters, one has the relation
\begin{equation}
 l_a\p_\xi f(l)=(\p_\xi l^b)L_{ab}f(l)\,,\qquad \text{or}\qquad
 \p_\xi f(l)= (t\cdot l)^{-1}t^a(\p_\xi l^b) L_{ab}f(l)\,,
\end{equation}
for any choice of $t$.
One of the often used parametrizations is given by $l=r(t+n)$, where $r>0$ and $n$ lies on the unit sphere in the hyperplane orthogonal to $t$, and may be parametrized by standard spherical angles $(\theta,\phi)$, or the related complex variables $(z,\bar{z})$, where $z=\tan(\theta/2)e^{i\phi}$.

\subsection{Null infinity}\label{ninf}

One of our main tasks will be the investigation of null asymptotes of fields. Suppose that for a given field $A(x)$, for each $x$ and $l\in C_+$, there exist limits
\begin{equation}
 \lim_{R\to\infty}RA(x\pm Rl)=a_\pm(x,l)\,.
\end{equation}
We want to make a motivated conjecture on the nature of the dependence of $a_\pm(x,l)$ on $x$. For this purpose let us stop the limit above at some large, but finite $R$. Let $x'$ lie on the past/future (for $+$/$-$, respectively)  light cone from the point $x\pm Rl$. Then we have $x\pm Rl=x'\pm R'l'$ for another $l'\in C_+$, and we scale this vector to $t\cdot l'=t\cdot l$. Then one finds: $R'=R+O(R^0)$, $l'=l+O(R^{-1})$, and for consistency vector $x'$ in finite separations from $x$ must satisfy $(x'-x)\cdot l=\pm(x'-x)^2/2R$, in the limit $x'\cdot l=x\cdot l$. This suggests that for $A$ sufficiently regular in null infinity, $a_\pm(x,l)$ are constant on null hyperplanes $x\cdot l=\con$. In that case we have
\begin{equation}\label{ninf_AV}
 \lim_{R\to\infty}RA(x+Rl)=V(x\cdot l,l)\,,\qquad
 \lim_{R\to\infty}RA(x-Rl)=V'(x\cdot l,l)\,,
\end{equation}
where $V(s,l)$, $V'(s,l)$ are functions on $\mR\times C_+$, with the property of homogeneity of degree $-1$: for $\la>0$ one has
\begin{equation}
 V(\la s, \la l)=\la^{-1}V(s,l)\,,\qquad
 V'(\la s, \la l)=\la^{-1}V'(s,l)\,;
\end{equation}
this follows easily if one scales $l\to\la l$ in relations \eqref{ninf_AV}.
Note that relations \eqref{ninf_AV} are Lorentz-covariant---independent of any choice of a reference basis or coordinates, and that both limits, for given $l$, are performed along the same null straight line. Suppose, moreover, that the limits $\dsp\lim_{R\to\infty}R\p_aA(x\pm Rl)$ exist, and are reached in uniform, locally in $x$, way. Then relations \eqref{ninf_AV} may be differentiated in $x$ and one finds
\begin{equation}\label{ninf_dAV}
 \lim_{R\to\infty}R\p_aA(x+Rl)=l_a\dV(x\cdot l,l)\,,\qquad
 \lim_{R\to\infty}R\p_aA(x-Rl)=l_a\dV'(x\cdot l,l)\,,
\end{equation}
where $\dV^\#(s,l)=\p V^\#(s,l)/\p s$. We shall see in the next subsection that fields to be considered indeed satisfy these relations.

We now specify relations \eqref{ninf_AV} to the special case when $x$ lies on a chosen time axis, denoted $x=ut$ for future expansion, or $x=vt$ for past expansion.  We scale $l$ to $t\cdot l=1$, and for this scaling write $R=r$. Then $u$ is the retarded time of the point $ut+rl$, and $v$ is the advanced time of the point $vt-rl$, and we find
\begin{equation}\label{ninf_Vu}
 \lim_{r\to\infty}rA(ut+rl)=V(u,l)\,, \quad
 \lim_{r\to\infty}rA(vt-rl)=V'(v,l)\,.
\end{equation}
To better understand the meaning of $V$ and $V'$, let us consider $A(ut+rl)$ and $A(vt-rl)$ with $r$ restricted to $r\in[0,r_0]$ for some constant $r_0$. Then arguments $ut+rl$, or $vt-rl$, cover a solid cylinder with radius $r_0$, and for each fixed $r$ and $l$,   parameters $u$ and $v$ are time parameters along a line parallel to the time axis. In particular, for $r=r_0$ the lines cover the boundary of the cylinder. For finite $r_0$ this boundary is common to both parametrizations, but when $r_0$ is taken to infinity, points on the boundary are propagated along null lines into future or past, in the two cases, the time span between limit future and past points becomes infinite, and the sets of limit values may differ.  In this interpretation $u$ in $V(u,l)$, and $v$ in $V'(v,l)$, are time parameters on different parts of the boundary cylinder of infinite radius. It is popular in recent literature to write expansions of $A(ut+rl)$ and $A(vt-rl)$ in powers of $1/r$. We write such expansions as
\begin{equation}\label{ninf_expr}
 A(ut+rl)=\sum_{n=1}\frac{1}{r^n}A^{+(n)}(u,z,\bar{z})\,,\quad
 A(vt-rl)=\sum_{n=1}\frac{1}{r^n}A^{-(n)}(v,z,\bar{z})\,.
\end{equation}
where $l=t+n$, $n$ is parametrized by variables $(z,\bar{z})$, as in Section \ref{lightcone}, and superscripts $\pm$ indicate future/past expansion. Then $A^{+(1)}(u,z,\zb)=V(u,l)$ and $A^{-(1)}(v,z,\zb)=V'(v,l)$. However, we note that the assumption on the existence of such expansions goes far beyond the existence of the limits \eqref{ninf_Vu} (see discussion at the end of Section 3.2 in \cite{her17}).

Another interpretation of \eqref{ninf_Vu} accounts better for the separation of limit regions. We eliminate $r$ in terms of retarded and advanced times,  $r=\frac{1}{2}(v-u)$, recall $l=t+n\in C_+$, and denote in addition
 $k=2t-l=t- n\in C_+$. It is then easy to see that relations \eqref{ninf_Vu} are equivalent to
\begin{equation}\label{ninf_Vuv}
 \lim_{v\to+\infty} \tfrac{1}{2}vA(\tfrac{1}{2}uk+\tfrac{1}{2}vl)=V(u,l)\,,\qquad
 \lim_{u\to-\infty} \tfrac{1}{2}|u|A(\tfrac{1}{2}vk+\tfrac{1}{2}ul)=V'(v,l)\,.
\end{equation}
Again, we restrict variables $u$, $v$ to $-T\leq u\leq v\leq T$ for some constant $T>0$. Then points $\tfrac{1}{2}uk+\tfrac{1}{2}vl$, or $\tfrac{1}{2}vk+\tfrac{1}{2}ul$, cover the diamond with vertices $-Tt, Tt$ (generally, a diamond with vertices $x_1, x_2$, where $x_2$ is in the future of $x_1$, is the intersection of the past solid light cone with vertex in $x_2$, with the future solid light cone with vertex in $x_1$). Consider the first equation, future limit, the past case is analogous. Points with fixed $v$ and $n$ form past light rays starting from $vt$, with $u\in[-T,v]$ an affine parameter along the null lines in direction $k$. In particular, for $v=T$, the lines $\tfrac{1}{2}uk+\tfrac{1}{2}Tl$, $u\in[-T,T]$, cover the future boundary of the region---part of the past light cone with vertex in $Tt$ . With $v$ replaced by $T$ in the first equation in \eqref{ninf_Vuv}, the vertex of this light cone goes to infinity along the time axis, and $V(u,l)$ is interpreted as the limit function on this large cone, with $u$ preserving the interpretation of an affine parameter along null rays along~$k$. In particular, if there exist limits of $V(u,l)$ for $u\to\pm\infty$, then $V(+\infty,l)$ is the limit value at the vertex in future infinity along $k$, while $V(-\infty,l)$ is the limit value at the 2-sphere in spacelike infinity, along $k$. In this interpretation, $V(u,l)$ may be seen as the lowest term coefficient in the expansion of $A(\tfrac{1}{2}uk+\tfrac{1}{2}vl)$ in the neighborhood of the future null infinity in powers of $(\tfrac{1}{2}v)^{-1}$. However, one should note that \eqref{ninf_expr} agrees with such expansion in the lowest term only, for higher orders the coefficients $A^{+(n)}(u,z,\bar{z})$ in \eqref{ninf_expr} are combinations of the coefficients of the latter expansion multiplied by powers of $u$.

The latter interpretation is often formalized by bringing infinity to a finite boundary by Penrose's method (see a textbook exposition in \cite{pen86}; see also \cite{str18} in the present context). This may be explained in two steps. First, variables $u,v$ are changed to $p,q$ by $u=\tan p$, $v=\tan q$, so that the spacetime is covered by the open diamond $-\pi/2<p\leq q<\pi/2$. One would like to close this diamond, adjoining the boundaries defined by subsets of $(p,q)$: future null infinity $\mathscr{I}^+= \{p\in(-\pi/2,\pi/2), q=\pi/2\}$, past null infinity $\mathscr{I}^-=\{p=-\pi/2, q\in(-\pi/2,\pi/2)\}$, one-point future timelike infinity \mbox{$i^+=\{p=q=\pi/2\}$}, one-point past timelike infinity $i^-=\{p=q=-\pi/2\}$, and spacelike infinity  $i^0=\{q=-p=\pi/2\}$, which is a $2$-sphere. Points $i^+$/$i^-$ are future/past boundaries, while the sphere $i^0$ is the past/future boundary, of $\mathscr{I}^+$/$\mathscr{I}^-$, respectively. However, on these boundaries the metric becomes infinite; therefore, in the next step one scales the metric conformally, multiplying it by factor $\cos^2(p)\cos^2(q)$, which  brings its form to $ds^2=4dpdq-\sin^2(q-p)(d\theta^2+\sin^2(\theta)d\phi^2)$. This produces a manifold with boundary, but this scaling changes the topology of the spacetime, as $i^0$ becomes one point (as $\sin(q-p)=0$ on $i^0$). One shows that this spacetime with boundary may be naturally interpreted as part of the `static Einstein universe' (product of a spacelike $3$-sphere with an infinite timelike line \cite{pen86}).

Conformal compactification helps to visualize null infinity asymptotics of solutions of wave equation (both homogeneous as inhomogeneous), and it may also be useful for technical reasons in the context of general asymptotically flat spacetimes. However, it is not convenient or helpful for the discussion of timelike asymptotics of massive matter or fields (which in this language originate/terminate in one point $i^-$/$i^+$, respectively), or spacelike asymptotics in general. An example of a different treatment of spacelike infinity is discussed in Section \ref{star} (and, in fact, already in formulas (\ref{Asp}, \ref{Fsp}) and (\ref{AAsp}, \ref{FFsp})).\footnote{And a different treatment of timelike infinity may be found in \cite{her95,her21}.} We shall not use the language of conformal compactification in these notes.

\subsection{Null and spacelike asymptotes in electrodynamics, matching property}\label{nullspace}

We review here the asymptotic properties of electrodynamics, as discussed in \cite{her95,her17}. We consider the Maxwell equations in Gauss' units:
\begin{equation}\label{max}
 \p_{[a}F_{bc]}(x)=0\,,\qquad \p_a F^{ab}(x)=4\pi j^b(x)\,,
\end{equation}
where $j^b$ is a conserved current: $\p\cdot j(x)=0$, which is a consistency condition for the second equation.
Solving the first equation one obtains $F_{ab}(x)=\p_aA_b(x)-\p_bA_a(x)$, with $A_a(x)$ the electromagnetic potential, determined up to a gauge transformation $A(x)\mapsto A(x)+\p\La(x)$, where $\La(x)$ is a scalar function. It is convenient to restrict the choice of potentials to the Lorenz class, defined by the second equation below, and then the first equation is equivalent to the second Maxwell equation:
\begin{equation}\label{AL}
 \Box A_a(x)=4\pi j_a(x)\,,\qquad \p\cdot A(x)=0\,.
\end{equation}
The Lorenz condition has the great advantages of being Lorentz-covariant, and reducing the dynamics to the inhomogeneous wave equation; in consequence, the space of dynamical variables is Lorentz invariant. Particular solutions of \eqref{AL} are retarded and advanced solutions, given by
\begin{equation}\label{retadv}
 A^{\ret/\adv}_a(x)=4\pi\int D^{\ret/\adv}(x-y)j_a(y)dy\,,\quad
 D^{\ret/\adv}(x)=\frac{1}{2\pi}\delta(x^2)\theta(\pm t\cdot x)\,,
\end{equation}
where $\delta$ is the Dirac delta and $\theta$ is the Heaviside step function. The current of a charged point-particle with charge $q$, moving along the world line $z^a(\tau)$, with $\tau$ its proper time and
$\dot{z}(\tau)=d z(\tau)/d\tau$ its local four-velocity, has the form
$j^a(x)=q\int \delta^{(4)}(x-z(\tau))\dot{z}^a(\tau)d\tau$. The ret/adv potentials of this current, the so called Lienard-Wiechert potentials, are often used in discussions of asymptotics, see e.g.\ \cite{str18}; we shall not need to consider these particular cases.

General solution of \eqref{AL} may be written in two alternative forms $A=A^\ret +A^\inc=A^\adv +A^\out$, where $A^{\inc/\out}$ satisfy homogeneous wave equation\footnote{In the following, by `wave equation' we shall always mean homogeneous wave equation, while the corresponding equation with source will be called `inhomogeneous wave equation'.} together with the Lorenz condition. These two forms may be considered as definitions of $\inc/\out$ fields, but their interpretation as incoming/outgoing fields has physical justification, see below.

Consider free fields first, and let $A^\fre$ stand for such field, in particular for $A^\inc$ or $A^\out$. Suppose that $A^\fre_a$ has asymptotes of the form \eqref{ninf_AV}, with $V^\fre_a(s,l)$ and $V^{\fre\prime}_a(s,l)$, and produces free electromagnetic field $F^\fre_{ab}$ with finite energy. It has been shown in \cite{her95} (see also \cite{her17}) that in this case there exist limits of asymptotic functions for $s\to\pm\infty$, and the following relations are satisfied
\begin{equation}\label{VVp}
 V^\fre_a(+\infty,l)=0=V^{\fre\prime}_a(-\infty,l)\,,\qquad
 V^\fre_a(s,l)+V^{\fre\prime}_a(s,l)
 =V^\fre_a(-\infty,l)=V^{\fre\prime}_a(+\infty,l)\,.
\end{equation}
The Lorenz condition is reflected by
\begin{equation}\label{VVL}
 l\cdot V^\fre(s,l)=l\cdot V^{\fre\prime}(s,l)=0\,.
\end{equation}
Moreover, if one denotes $\dV^\fre(s,l)=\p_sV^\fre(s,l)$, then the potential is recovered in the whole spacetime by formula
\begin{equation}\label{AVint}
 A^\fre_a(x)=-\frac{1}{2\pi}\int \dV^\fre_a(x\cdot l,l)\,d^2l\,,
\end{equation}
in which one could equivalently use $\dV^{\fre\prime}_a$ in place of $-\dV^\fre_a$.
This formula may be interpreted as a special case of the Kirchhoff formula for recovering solution of the wave equation from its data on a null hypersurface\footnote{Standard initial value problem for a solution of the wave equation requires knowledge of its restriction to a spacelike Cauchy hypersurface, and the restriction of its derivative in the direction orthogonal to the hypersurface. When the hypersurface is deformed to a null hypersurface, the direction orthogonal lies in this hypersurface, which explains why the initial values reduce to one function in that case.} (see, e.g.\ \cite{pen84}), in this case future null infinity (see \cite{her98} for details). Conversely, let $A^\fre(x)$ have that form, suppose that $|\dV^\fre(s,l)|\leq\con(1+|s|)^{-1-\vep}$, $\vep>0$, and choose $V^\fre(s,l)$ with $V(+\infty,l)=0$ (this is a slight strengthening of the assumption on the  limits of $V^\fre(s,l)$ for $s\to\pm\infty$). Then $A^\fre$ has asymptotes $V^\fre$ and $V^{\fre\prime}$ satisfying relations \eqref{VVp}, \eqref{VVL} and \eqref{AVint}. We would like to stress an important point: formula \eqref{AVint} depends on $\dV^\fre(s,l)$, so it would seem the function $V^\fre(s,l)$ is determined up to the addition of an $s$-independent term. However, for $V^\fre(s,l)$ to be the future asymptote one has to satisfy $V^\fre(+\infty,l)=0$, so the function is uniquely fixed. Similarly for $V^{\fre\prime}$. The corresponding free field has the form
\begin{equation}\label{FVint}
 F^\fre_{ab}(x)=-\frac{1}{2\pi}\int \big[l_a\ddV^\fre_b(x\cdot l,l)-l_b\ddV^\fre_a(x\cdot l,l)\big]d^2l\,.
\end{equation}

We now note that the difference $ A^\out-A^\inc$ is equal to $A^\rad = A^\ret-A^\adv$, the so called radiation potential of the current $j$, so for the consistency of the picture this field should also satisfy relations \eqref{VVp}, \eqref{VVL} and \eqref{AVint}. Using formulas \eqref{retadv} one finds
\begin{equation}\label{radpot}
 A^\rad(x)=4\pi\int D(x-y)j(y)dy\,,\quad
 D(x)=\frac{1}{2\pi}\delta(x^2)\sgn(x\cdot t)=-\frac{1}{8\pi^2}\int \delta'(x\cdot l)\,d^2l\,,
\end{equation}
where $D(x)$ is the so called Pauli-Jordan function, and its rhs form is easily proved by performing integration ($\delta'$ is the derivative of the Dirac delta function). Using this form in the integral for $A^\rad$ one finds that
\begin{equation}\label{radpotV}
  A^\rad(x)=-\frac{1}{2\pi}\int \dV^j(x\cdot l,l)\,d^2l\,,
\end{equation}
where
\begin{equation}\label{Vj}
 V^j_a(s,l)=\int \delta(s-x\cdot l)j_a(x)dx\,,\qquad l\cdot V^j(s,l)=q\,,
\end{equation}
where $q$ is the electric charge of the current $j$.
Now, for typical currents of a scattering setting this function is regular (differentiable of some positive degree), and has limits for $s\to\pm\infty$, which is related to the fact, that the dominant form of the current $j(x)$ in far past and future is supported inside the light cone and is a homogeneous function of degree $-3$. Assuming, as before, a slightly stronger condition $|\dV^j(s,l)|\leq\con(1+|s|)^{-1-\vep}$, one finds that $A^\rad$ has asymptotes \eqref{ninf_AV} $V^\rad$ and $V^{\rad\prime}$ given by
\begin{equation}
 V^\rad(s,l)=V^j(s,l)-V^j(+\infty,l)\,,\qquad
 V^{\rad\prime}(s,l)=-V^j(s,l)+V^j(-\infty,l)\,,
\end{equation}
so the consistency condition mentioned earlier is satisfied.\footnote{One could reverse the direction of the argument: start with the stated properties of current $j$, build its retarded and advanced fields, and consider its radiation potential \eqref{radpot}. Then it is natural to assume that `in' potential is also of \eqref{AVint} type, and consequently the same is then true for `out'.}

Next, we consider null asymptotes of the retarded and advanced fields. One shows that
\begin{gather}
 V^\ret(s,l)=V^j(s,l)\,,\qquad V^{\ret\prime}(s,l)=V^j(-\infty,l)\,,\label{retas}\\
 V^\adv(s,l)=V^j(+\infty,l)\,,\qquad V^{\adv\prime}(s,l)=V^j(s,l)\,.\label{advas}
\end{gather}
Formulas for $V^\ret$ and $V^{\adv\prime}$ are rather easily obtained with the use of \eqref{retadv}, while the other two formulas follow most easily from identities
$V^\adv=V^\ret-V^\rad$, $V^{\ret\prime}=V^{\adv\prime}+V^{\rad\prime}$.

There is one additional property of asymptotes in standard electromagnetic settings, related to the fact that fields are produced by electric charges, and there are no magnetic charges in the theory. This implies
\begin{equation}\label{antL}
 L_{[ab}V_{c]}^\#(\pm\infty,l)=0\,,
\end{equation}
where $V^\#$ is any of the considered functions.

Collecting results on asymptotes of $A^{\ret/\adv}$ and $A^{\inc/\out}$, one finds that the total potential $A$ has asymptotes $V(s,l)$ and $V'(s,l)$, satisfying relations
\begin{gather}
  V_a(s,l)+V'_a(s,l)-V_a^j(s,l)=V_a(-\infty,l)=V'_a(+\infty,l)\,,\label{match}\\
 V_a(+\infty,l)=V_a^j(+\infty,l)\,,\qquad
 V'_a(-\infty,l)=V_a^j(-\infty,l)\,,\label{VVj}\\
 L_{[ab}V_{c]}(\pm\infty,l)=L_{[ab}V'_{c]}(\pm\infty,l)=0\,,\qquad
  l\cdot V(s,l)=l\cdot V'(s,l)=q\,,\label{LV}\\
  V_a^\out(s,l)=V_a(s,l)-V_a(+\infty,l)\,,\quad
 V_a^{\inc\prime}(s,l)=V_a'(s,l)-V_a'(-\infty,l)\,.\label{VVVV}
\end{gather}
The corresponding asymptotes for the field $F_{ab}$ are obtained by the application of \eqref{ninf_dAV}. Thus we can summarize explicit asymptotic formulas
\begin{gather}
 \lim_{R\to\infty}RA_a(x+Rl)=V_a(x\cdot l,l)\,,\quad
 \lim_{R\to\infty}RF_{ab}(x+Rl)
 =l_a\dV_b(x\cdot l,l)-l_b\dV_a(x\cdot l,l)\,,\label{nullfut}\\
 \lim_{R\to\infty}RA_a(x-Rl)=V'_a(x\cdot l,l)\,,\quad
 \lim_{R\to\infty}RF_{ab}(x-Rl)
 =l_a\dV'_b(x\cdot l,l)-l_b\dV'_a(x\cdot l,l)\,.\label{nullpast}
\end{gather}

The second equality in \eqref{match} is the `matching property', which was obtained in \cite{her95} (eqn.\ (2.26)), and whose free field version could have already been noticed in the last equation of \eqref{VVp}. Note that this matching equates limit values on the $2$-sphere in spacelike infinity in antipodal positions, which is the reason for the name of `antipodal matching' used recently for identities of this type. However, in our opinion more significant and explicative is the fact that the values are reached along the same null direction $l$. Note also, that the matching property is a consequence of the existence of future and past null asymptotes, and not an independent `matching condition', as it is often recently called.

The (equal) variables on the rhs of \eqref{match} govern the asymptotic spacelike behavior of the fields \cite{her95,her17}: for spacelike vectors $y$, independent of the choice of a point $x$, one obtains asymptotic limit functions\footnote{In Ref.\ \cite{her17}, factors $1/(2\pi)$ are lacking.}
\begin{align}
 A^\as_a(y)&=\lim_{R\to\infty}RA_a(x+Ry)
  =\frac{1}{2\pi}\int V_a(-\infty,l)\,\delta(y\cdot l)\,d^2l\,,\label{Asp}\\
 F^\as_{ab}(y)&=\lim_{R\to\infty}R^2F_{ab}(x+Ry)
  =\frac{1}{2\pi}
  \int\big[l_a V_b(-\infty,l)-l_b V_a(-\infty,l)\big]
  \delta'(y\cdot l)\,d^2l\,,\label{Fsp}
\end{align}
where $\delta$ and $\delta'$ are the Dirac delta and its derivative, respectively.  The rhs of these relations do not depend on the spacetime point $x$, so they may be interpreted as conserved quantities. The asymptotic potential and electromagnetic field are homogeneous functions of degree $-1$ and $-2$, respectively, which shows that fields considered in this section decay in spacelike infinity at Coulomb-like rate. One can show that for free fields, when $l\cdot V(s,l)=0$, the above limits also hold distributionally with $y\in M$ (although not pointwise for $y^2=0$), yielding a global solution $F^\as_{ab}(y)$ of the free Maxwell equations. For $l\cdot V(s,l)=q$ the limits do not hold in this form for $y^2\geq0$, but still the formula for $F^\as_{ab}(y)$ given by the rhs may be extended to the whole space, yielding a distributional field satisfying $\p^aF^\as_{ab}(y)=4\pi q\p_bD(y)$ with the Pauli-Jordan function $D(y)$ \eqref{radpot} on the rhs. One can easily show that the charge of the distributional current $q\p_bD(y)$ is, indeed, equal to $q$ (although, the current itself, carrying charge expanding in lightlike fashion, does not seem physically realistic).

\subsection{Gauge transformations, problem of LGT, general fields of Coulomb-like decay}\label{gauge}

Let $A$ and $A^\Lambda=A+\p\La$ be two potentials from the class of Section \ref{nullspace}. The Lorentz condition implies then $\Box\La(x)=0$, but more than that, $\p\La$ is a free Lorenz potential in our class, so the representation \eqref{AVint} gives
\begin{equation}\label{La}
 \p_a\La(x)=-\frac{1}{2\pi}\int \dW_a(x\cdot l,l)\,d^2l\,,\qquad \text{and}\qquad
 0=\p_{[a}\p_{b]}\La(x)
 =-\frac{1}{2\pi}\int l_{[a}\ddot{W}_{b]}(x\cdot l,l)\,d^2l
\end{equation}
for some $\dW_a(s,l)$ in our class. Taking the future null asymptote of the zero field we find $l_{[a}\dW_{b]}(s,l)=0$, so $\dW_a(s,l)=l_a\dot\alpha(s,l)$. But $W_a(s,l)$ has limits in $s=\pm\infty$, equal $0$ for $s=+\infty$, so the same is true for $\al(s,l)$, and $W_a(s,l)=l_a\al(s,l)$. Substituting $\dW_a(x\cdot l,l)=l_a\dot\al(x\cdot l,l)$ into the first relation in \eqref{La}, we finally can summarize. Gauge transformations in our class are given by
\begin{equation}
 A_a(x)\to A_a(x)+\p_a\La(x)\,,\qquad
 \La(x)=-\frac{1}{2\pi}\int \al(x\cdot l,l)\,d^2l +\al_+\,,
\end{equation}
where $\al_+$ is a numerical constant. If we now substitute $x\to x\pm Rl$ (changing the integration vector to $l'$), take into account that $\alpha(x\cdot l\pm Rl\cdot l',l')$ is bounded and for $R\to\infty$ tends almost everywhere in $l'$ to $0$ for $+$, and to $\alpha(-\infty,l)$ for $-$, we find
\begin{equation}\label{limLa}
 \lim_{R\to\infty} \La(x+Rl)=\alpha_+\,,\qquad
 \lim_{R\to\infty}\La(x-Rl)=\alpha_-
 =\alpha_+ -\frac{1}{2\pi}\int \alpha(-\infty,l')\,d^2l'\,.
\end{equation}
Thus the asymptotic limits of gauge functions respecting the demands of the formalism are constant over the future and the past null infinities, with generically different constants.

More recently, another type of transformation, named `large gauge transformation' (LGT), has been proposed (see \cite{cam15}, \cite{kap17}, \cite{str18} and references in these articles). This problem may be approached in two ways. First, one can pose a question: are there other gauges, obtained from the above Lorenz type by a transformation $A_a(x)\to A^\La_a(x)=A_a(x)+\p_a\Lambda(x)$, such that $A^\La_a(x+Rl)$ has again a null asymptote of order $R^{-1}$? One shows (this is discussed in details in \cite{her17}, Section 7) that if $\Lambda(st+Rl)=\vep^+(l)+R^{-1}\beta_t(s,l)+o(R^{-1})$, where we have scaled $t\cdot l=1$ for simplicity, and $\beta_t$ is appropriately correlated with $\vep^+$, then
\begin{equation}\label{VVinf}
 \lim_{R\to\infty}RA^\La_a(st+Rl)=V^\La_a(s,l)=V_a(s,l)+V_a^+(l)\,,
\end{equation}
where $V_a^+(l)$ and $\vep^+(l)$ are related as $V(l)$ and $\phi(l)$ in (\ref{fsp}--\ref{Vfi}). We analyze this change from the point of view, and in the language of, the future null infinity of Section \ref{nullspace}. In terms of the outgoing and the advanced fields we have
\begin{equation}
 V^{\La\out}(s,l)=V^\La(s,l)-V^\La(+\infty,l)=V^\out(s,l)\,,\quad
 V^{\La\adv}(s,l)=V^\La(+\infty,l)=V(+\infty,l)+V^+(l)\,.
\end{equation}
Therefore, the potential of the outgoing field remains unchanged. On the other hand, recall that $V(+\infty,l)=V^j(+\infty,l)$, so the addition $V^+(l)$ has the interpretation of the potential contribution due to an additional outgoing electromagnetic current (charge-free in this case, as $l\cdot V^+(l)=0$). Note also that $V^\La(-\infty,l)=V(-\infty,l)+V^+(l)$, and $l_aV_b^+(l)-l_bV^+_a(l)\neq0$, so this addition is not a gauge change within this formalism, as the spacelike asymptote \eqref{Fsp} clearly shows.

To better understand this contradiction let us approach the problem from  another side. In the search for a description of LGT one would like to have a gauge function with $l$-dependent limits $\al_\pm(l)$, instead of constants $\al_\pm$ as in \eqref{limLa}. And indeed, as proposed in \cite{cam15}, and reviewed in \cite{str18}, the following function
\begin{equation}
 \la(x)
 =-\frac{1}{4\pi}\int
 \bigg\{\log\Big|\frac{x\cdot l}{t\cdot l}\Big|\p^2\al(l)-\frac{\al(l)}{(t\cdot l)^2}\bigg\}\,d^2l
 =\frac{x^2}{4\pi}\int \frac{\al(l)}{(x\cdot l)^2}\,d^2l
\end{equation}
(the second equality by \eqref{pV0}) satisfies wave equation and has limits
\begin{equation}\label{limla}
 \lim_{R\to\infty} \la(x\pm Rl)=\al(l)\,.
\end{equation}
However, the corresponding gauge contribution to the potential,
\begin{equation}\label{potla}
 \p_a\la(x)=-\frac{1}{4\pi}\int \frac{l_a\p^2\al(l)}{x\cdot l}\,d^2l\,,
\end{equation}
is not in the class of potentials \eqref{AVint}, as it corresponds to the function $\dV^\la_a(s,l)=\tfrac{1}{2}l_a\p^2\al(l)/s$.  And indeed, it does not have the null infinity behavior assumed in the formalism of Section \ref{nullspace}. One can show that for $R\to\infty$ one has
\begin{equation}\label{aspla}
 \p_a\la(x\pm Rl)=\mp\frac{\log R}{R}\,\tfrac{1}{2}l_a\p^2\al(l)+O(R^{-1})\,.
\end{equation}
Functions $\la$ and $\p\la$ are not regular on the cone, and in addition the rest on the rhs of \eqref{aspla} is singular on $x\cdot l=0$, but these are artefacts due to the singularity of $\dV^\la(s,l)$ in $s=0$; any smooth function $V(s,l)$, even in $s$, with $\dV(s,l)$ approaching (sufficiently fast) $\dV^\la(s,l)$ for large $|s|$, will produce smooth function with unchanged limits \eqref{limla} and asymptotic behavior \eqref{aspla}.

This needs to be placed in a wider context. Note that formula \eqref{AVint} defines a Lorenz potential of a free field for each smooth (or at least sufficiently regular) function $\dV^\fre(s,l)$, orthogonal to $l$, not necessarily submitted to decay conditions. Let us denote the Fourier transform
\begin{equation}\label{Fou}
 \ti{\dV^\fre_a}(\w,l)=\frac{1}{2\pi}\int \dV^\fre_a(s,l)e^{i\w s}ds\,,\qquad
 \ov{\ti{\dV^\fre_a}(\w,l)}=\ti{\dV^\fre_a}(-\w,l)\,,
\end{equation}
where the second formula follows from reality of $V^\fre(s,l)$. One shows that the existence of the limit $\dsp \ti{\dV^\fre_a}(0^+,l)=\lim_{\w\searrow0}\ti{\dV^\fre_a}(\w,l)$ is equivalent to the general Coulomb rate decay in spacelike infinity of the potential \eqref{AVint}, and its field. For $y^2<0$ one finds
\begin{align}
 A^{\fre\as}_a(y)&=\lim_{R\to\infty}RA^\fre_a(x+Ry)
  =\frac{i}{2\pi}\int
  \frac{\ti{\dV^\fre_a}(0^+,l)}{y\cdot l-i0}\,d^2l
  +\text{compl. conj.}\,,\label{AAsp}\\
 F^{\fre\as}_{ab}(y)&=\lim_{R\to\infty}R^2F^\fre_{ab}(x+Ry)
  =-\frac{i}{2\pi}
  \int\frac{l_a \ti{\dV^\fre_b}(0^+,l)-l_b\ti{\dV^\fre_a}(0^+,l)}{(y\cdot l-i0)^2}\,d^2l+\text{compl. conj.}\,.\label{FFsp}
\end{align}
If we assume in addition that $L_{[ab}\ti{\dV^\fre_{c]}}(0^+,l)=0$ then we obtain the most general class of fields $F^\fre_{ab}$ with electric, Coulomb-like decay in spacelike infinity. Similarly as for free fields in  (\ref{Asp}, \ref{Fsp}), these more general asymptotic formulas may be extended to all $y$, yielding global distributional solutions of the free Maxwell equations. From the second equality in \eqref{Fou} we have $\dsp\ov{\ti{\dV^\fre}(0^+,l)}=\ti{\dV^\fre}(0^-,l)$, so the real part of $\ti{\dV^\fre}(\w,l)$ is continuous in $\w=0$, while imaginary part may have a jump. The subclass with $\Ip{\ti{\dV^\fre}(0^+,l)}=0$ is our class of Section \ref{nullspace}, and the above asymptotic formulas reproduce then free field case of \eqref{Asp} and \eqref{Fsp}. However, if $\Ip{\ti{\dV^\fre}(0^+,l)}\neq0$, then the jump in $\w=0$ produces $V^\fre(s,l)$ with dominant behavior of the form
\begin{equation}\label{gc}
 \dV^\fre(s,l)=\frac{2}{s}\Ip{\ti{\dV^\fre}(0^+,l)}+ O(|s|^{-1-\vep})\qquad
 \text{for}\qquad |s|\to\infty\,,
\end{equation}
and the dominant null asymptotic behavior of potentials is of type \eqref{aspla}. Therefore, a more proper setting for the analysis of this larger set is spacelike infinity. It is now clear, that $\p\la(x)$ \eqref{potla} is the spacelike asymptote of a trivial (pure gauge) potential in this larger class, and its proper place is this general class.

It is worth noting that the asymptotic fields \eqref{Fsp} are odd functions of $y$, while the additional fields in \eqref{FFsp} are even.
One should also stress, that fields with Coulomb-like, even in $y$ decay, do not appear in usual physical settings. For instance, to produce such field as a radiation field of a current source (with potential as in (\ref{radpot}, \ref{Vj})), one would have to use $j(x)$ decaying in timelike directions as $\log|x|/|x|^3$, as opposed to the decay $1/|x|^3$ in usual scattering settings.\footnote{The field tensors $F_{ab}$ in that class do have null asymptotes, but neither the symplectic form in null infinity to be discussed in Section \ref{sympstr}, nor such physical quantity as angular momentum radiated into infinity (even over finite time spans, \cite{her95}), are well defined.}

We refer the reader to Section \ref{star}, where we review Staruszkiewicz's theory of asymptotic fields \eqref{Fsp} and \eqref{FFsp}, and a certain phase function $S(x)$. Also, we note that the above function $\la(x)$, as a homogeneous, even function, is identical with part of $S(x)$.

\subsection{Matching once more, and charges}\label{matchar}

One can smear $V(-\infty,l)$ with a test vector function $V^+(l)$, homogeneous of degree $-1$ and orthogonal to~$l$, to obtain a numerical quantity
\begin{equation}\label{charge}
 Q(V^+)=-\frac{1}{4\pi}\int V^+(l)\cdot V(-\infty, l)\,d^2l\,,
\end{equation}
which does not depend on the gauge of $V(-\infty,l)$. We shall find an alternative form of this quantity. Making use of \eqref{Vfp}, we represent
$l_aV^+_b(l)-l_bV^+_a(l)=L_{ab}\vep^+(l)-{}^*\!L_{ab}\kappa^+(l)$, and contract this equation with $t^aV^b(-\infty,l)$, with arbitrarily chosen $t$. In this way we find
\begin{equation}
 V^+(l)\cdot V(-\infty,l)
 =t^a\big[L_{ab}\vep^+(l)-{}^*\!L_{ab}\kappa^+(l)\big]
 \frac{V^b(-\infty,l)}{t\cdot l}
 +q\frac{t\cdot V^+(l)}{t\cdot l}\,,
\end{equation}
where we have taken into account that $l\cdot V(-\infty,l)=q$. When this identity is integrated, operators $L$ and ${}^*\!L$ are transferred by \eqref{Lf0} to $(t\cdot l)^{-1}V^b(-\infty,l)$. But ${}^*\!L_{ab}V^b(-\infty,l)=0$ due to \eqref{antL}, and also $t^a{}^*\!L_{ab} t\cdot l=0$. Therefore, without restricting generality we assume that $\kappa^+(l)=0$ for $V^+(l)$.
We represent function $\vep^+(l)$ as in \eqref{fsp}, and note that then for the pair $\vep^+(l)$, $V^+(l)$ relation \eqref{fspid} is satisfied. Taking this into account one finds
\begin{equation}\label{chargee}
 Q(V^+)=\frac{1}{4\pi}\int \vep^+(l)\p\cdot V(-\infty,l)\,d^2l\,,
\end{equation}
where it has been assumed that for the sake of differentiation $V(-\infty,l)$ is extended to a neighborhood of the cone as homogeneous of degree $-1$ and satisfying $l\cdot V(-\infty,l)=q$.

We want now to compare this with results formulated more recently in the language of future null expansions of the form \eqref{ninf_expr}. When tensor quantities are thus expanded, the expansions of this form apply to their algebraic basis coordinates. However, authors usually use coordinate basis of coordinates $u,r,z,\bar{z}$ for future, and $v,r,z,\zb$ for past. As  $x=ut+rl$ and $x=vt-rl$ in these two coordinations, with $l=t+n$, the coordinate basis is
 $(\p_ux, \p_rx, \p_zx, \p_{\zb}x)=(t, l, r\p_zn, r\p_{\zb}n)$ in the retarded coordinates, and
 $(\p_vx, \p_rx, \p_zx, \p_{\zb}x)=(t, -l, -r\p_zn, -r\p_{\zb}n)$ in the advanced coordinates. Therefore,
various components get additional $r$-factors, which is a complication to be taken into account. For instance, for the potential $A_a(x)$ expanded as in \eqref{ninf_expr} one has
\begin{gather}
\begin{gathered}
 A^{+(n)}_u(u,z,\zb)=t^aA^{+(n)}_a(u,z,\zb)\,,\quad  A^{+(n)}_r(u,z,\zb)=l^aA^{+(n)}_a(u,z,\zb)\,,\\
 A^{+(n)}_z(u,z,\zb)=\p_zn^aA_a^{+(n+1)}(u,z,\zb)\,,
\end{gathered}\\
\begin{gathered}
 A^{-(n)}_v(v,z,\zb)=t^aA^{(n)}_a(v,z,\zb)\,,\quad  A^{-(n)}_r(v,z,\zb)=-l^aA^{-(n)}_a(v,z,\zb)\,,\\
 A^{-(n)}_z(v,z,\zb)=-\p_zn^aA_a^{-(n+1)}(v,z,\zb)\,,
\end{gathered}
\end{gather}
(the $\zb$-components are conjugates of $z$-components) and similarly for higher rank tensors. However, for the $r,u$- or $r,v$-component of the electromagnetic field one has $F_{ru}=l^at^bF_{ab}$ or $F_{rv}=-l^at^bF_{ab}$, respectively, so the orders agree and one has
\begin{equation}
 F^{+(n)}_{ru}(u,z,\zb)=l^at^bF^{+(n)}_{ab}(u,z,\zb)\,,\quad
 F^{-(n)}_{rv}(v,z,\zb)=-l^at^bF^{-(n)}_{ab}(v,z,\zb)\,,
\end{equation}
The asymptotes of $F_{ab}$ given in \eqref{nullfut} and \eqref{nullpast} vanish, when contracted with $l^a$, so the lowest terms with $n=1$ above are zero. We need the next to leading terms now.
We form an auxiliary field $B_b(x)=x^aF_{ab}(x)$ and note that its expansions follow from those of $F_{ab}(x)$. In particular, the lowest terms are related by
 $t^bB^{+(1)}_b(u,z,\zb)=F^{+(2)}_{ru}(u,z,\zb)$ and $t^bB^{-(1)}_b(v,z,\zb)=F^{-(2)}_{rv}(v,z,\zb)$.
It is thus sufficient to calculate the asymptotes
\begin{equation}
 \lim_{R\to\infty}RB_b(x+Rl)=W_b(x\cdot l,l)\,,\quad
 \lim_{R\to\infty}RB_b(x-Rl)=W'_b(x\cdot l,l)\,,
\end{equation}
which are shown to be given by
\begin{equation}
 W_b(s,l)=L_{ba}V^a(s,l)-V_b(s,l)+s\dV_b(s,l)\,,\quad
 W'_b(s,l)=L_{ba}V^{\prime a}(s,l)-V'_b(s,l)+s\dV'_b(s,l)\,.
\end{equation}
We sketch a proof for future infinity. We have
\begin{equation}
 RB_b(x+Rl)=Rx^aF_{ab}(x+Rl)
 + R^2l\cdot \p A_b(x+Rl)-R^2(l_a\p_b-l_b\p_a)A^a(x+Rl)\,.
\end{equation}
The asymptote of the first term follows from the asymptote of $F_{ab}$, and we obtain in the limit
 \mbox{$x\cdot l\dV_b(x\cdot l,l)-l_bx\cdot \dV(x\cdot l,l)$}. The second and
the third terms may be written as $R^2\p_RA_b(x+Rl)$ and $RL_{ba}A^a(x+Rl)$, respectively, which gives in the limit
$-V_b(x\cdot l,l)$ and $L_{ba}V^a(x\cdot l,l)$, respectively. Performing differentiation in the last term one obtains
 $L_{ba}V^a(s,l)|_{s=x\cdot l}+l_bx\cdot \dV(x\cdot l,l)$. Adding up the terms
one obtains $W_b(x\cdot l,l)$. Similarly for the past asymptote.

We now have $F^{+(2)}_{ru}(u,z,\zb)=t\cdot W(u,t+n)$, $F^{-(2)}_{rv}(v,z,\zb)=t\cdot W'(v,t+n)$. Now, for $B$ similar matching as for $A$ is satisfied: $W_b(-\infty,l)=W'_b(+\infty,l)$, which in fact is already a consequence of the matching \eqref{match}. Contracted with $t^b$ it gives
\begin{equation}\label{strm}
 F^{+(2)}_{ru}(-\infty,z,\zb)=F^{-(2)}_{rv}(+\infty,z,\zb)\,,
\end{equation}
which is the matching discussed in \cite{str18} (and, in fact, obtained there  only on the basis of examples, including Lienard-Wiechert potentials). Now, if $V(-\infty,l)$ and $V'(+\infty,l)$ are extended into a neighborhood of the cone with the preservation of homogeneity of degree $-1$ and of the condition $l\cdot V(-\infty,l)=l\cdot V'(+\infty,l)=q$, then one finds  $W_a(-\infty,l)=l_a\p\cdot V(-\infty,l)$ and similarly for $W'$. Thus the invariant formulation of Strominger's matching takes now the form
\begin{equation}
 \p\cdot V(-\infty,l)=\p\cdot V'(+\infty,l)\,,
\end{equation}
an obvious consequence of \eqref{match}.
When these quantities are smeared with $\vep^+(l)/4\pi$, one obtains $Q(V^+)$ in explicit form \eqref{chargee}. On the other hand, when $l$'s are scaled to $t\cdot l=1$ and integration $d^2l$ is specified to $d\W(\lb)$, this becomes identical with smeared relation \eqref{strm}, giving `charges' in Strominger's terminology.

There is one more interesting smearing property related to the form of field  in spacelike infinity. We note that $B_b(x)$ has the spacelike asymptote similar to that for $A_b(x)$ given in \eqref{Asp}, with $V_b(-\infty,l)$ under the integral replaced by $l_a\p\cdot V(-\infty,l)$. Therefore, we obtain
\begin{equation}
 \lim_{R\to\infty}R^2y^aF_{ab}(x+Ry)=\lim_{R\to\infty}RB_b(x+Ry)
 =\frac{1}{2\pi}\int l_b\p\cdot V(-\infty,l)\delta(y\cdot l)\,d^2l\,.
\end{equation}
Let us contract this with $t^b$, multiply by a test function $\varphi(y)$   with compact support in $y^2<0$, integrate with $dy$, and denote $\vep^+(l)=t\cdot l\int \varphi(y)\delta(y\cdot l)dy$, which is a homogeneous function. Then such average of the field in spacelike infinity gives precisely the quantity \eqref{chargee}.

Having described all necessary asymptotic properties of the formalism, we go back once more to a free potential $A^\fre$ and field $F^\fre$. The integral representation \eqref{AVint} of a free field potential is related to the more standard Fourier representation
\begin{equation}\label{Aint}
 A^\fre_a(x)
 =\frac{1}{\pi}\int e^{-ix\cdot k}a^\fre_a(k)\sgn(k^0)\delta(k^2)d^4k
\end{equation}
by
\begin{equation}\label{aV}
 \w a^\fre_a(\w l)=-\ti{\dV^\fre_a}(\w,l)
 =-\frac{1}{2\pi}\int e^{i\w s}\dV^\fre_a(s,l)ds\,,
\end{equation}
so one has a sequence of identities
\begin{equation}\label{aV0}
\begin{aligned}
 \lim_{\w\to0}\w a^\fre_a(\w l)
 =-\ti{\dV^\fre_a}(0,l)
 &=\frac{1}{2\pi}V^\fre_a(-\infty,l)
 =-\frac{1}{2\pi}\int \dV^\fre_a(s,l)ds\,,\\
 =\ti{\dV^{\fre\prime}_a}(0,l)
 &=\frac{1}{2\pi}V^{\fre\prime}_a(+\infty,l)
 =\frac{1}{2\pi}\int \dV^{\fre\prime}_a(s,l)ds\,.
\end{aligned}
\end{equation}
Note that these equations say, in particular, that $\w a^\fre(\w l)$ is continuous at $\w=0$, in agreement with our discussion at the end of Section \ref{gauge}.

Looking at equations (\ref{Asp}, \ref{Fsp}), applied to our free field, we see that if any of the equal quantities \eqref{aV0} does not vanish, then  $F^\fre$ is infrared singular. For the incoming and the outgoing fields one thus has
\begin{equation}\label{aVoutin}
\begin{aligned}
 \lim_{\w\to0}\w a^\out_a(\w l)
 &=\frac{1}{2\pi}V^\out_a(-\infty,l)
 =\frac{1}{2\pi}[V_a(-\infty,l)-V_a(+\infty,l)]
 =-\frac{1}{2\pi}\int \dV_a(s,l)ds\,,\\
 \lim_{\w\to0}\w a^\inc_a(\w l)
 &=\frac{1}{2\pi}V^{\inc\prime}_a(+\infty,l)
 =\frac{1}{2\pi}[V'_a(+\infty,l)-V'_a(-\infty,l)]
 =\frac{1}{2\pi}\int \dV'_a(s,l)ds\,.
\end{aligned}
\end{equation}
These identities, a simple consequence of \eqref{aV0} and \eqref{VVVV}, relate the traditional infrared characteristics of free fields on the lhs, to the quantities on the edges of future null infinity and to the integral of the electromagnetic field asymptote.

We end this section with an illustration of the `matching
property' \eqref{match}, as described in \cite{her17}. Consider a scattering event in which there are $n'$ incoming particles and $n$ outgoing particles,
with charges and four-velocities given by $q'_i, v'_i$ and $q_i,
v_i$, respectively. It is easy to show that for the current $j(x)$ due to a single free
particle with charge $q$ and velocity $v$ one has
$V^j(s,l)=q\,v/v\cdot l$. Taking also into account relations \eqref{VVj}, in our scattering event we
have
\begin{equation}
 V'(-\infty,l)=V^j(-\infty,l)
 =\sum_{i=1}^{n'}q'_i\,\frac{v'_i}{v'_i\cdot l}\,,\quad
 V(+\infty,l)=V^j(+\infty,l)=\sum_{i=1}^{n}q_i\,\frac{v_i}{v_i\cdot l}\,.
\end{equation}
Then relations \eqref{match} and \eqref{aV0} give
\begin{equation}\label{infra}
 2\pi \lim_{\w\to0}\w a^\out(\w l)
 +\sum_{i=1}^{n}q_i\frac{v_i}{v_i\cdot l}=V(-\infty,l)=V'(+\infty,l)
 =2\pi\lim_{\w\to0}\w a^\inc(\w l)
 +\sum_{i=1}^{n'}q'_i\frac{v'_i}{v'_i\cdot l}\,.
\end{equation}
It is now clear that unless the scattering of charges is trivial (i.e.\ $n=n'$, $q_i=q'_i$ and $v_i=v_i'$), the incoming and the outgoing fields cannot be simultaneously infrared regular, which leads to problems in standard formulations of QED, as we shall see below. Equality of outer sides in \eqref{infra} may be viewed as a classical prototype of what in quantum world has a consequence of so called soft photon theorems, see below.

\subsection{Electromagnetic memory effects}\label{mem}

An effect of the type of `memory', first discovered for gravitation \cite{zel74,bra85}, has been first noted in electrodynamics by Staruszkiewicz \cite{sta81}, who posed the following question: does a (free)  electromagnetic field in zero frequency limit produce observable effects? What is meant by zero frequency is the following. Let $A^\fre_a(x)$ be a Lorenz potential of a field from the class identified in Section \ref{nullspace}, and define its rescaled version $A_a^{\fre(\la)}(x)=\la^{-1}A^\fre_a(\la^{-1}x)$, $\la>0$, which is equivalent to $\dV^{\fre(\la)}(s,l)=\la^{-1}\dV^\fre(\la^{-1}s,l)$. Consider the scaled field in the large $\la$ limit. The energy carried by this field vanishes in that limit, so it is unable to change the velocity of any massive charged particle. In terms of the Fourier transform, $\ti{\dV_a^{\fre(\la)}}(\w,l)=\ti{\dV^\fre_a}(\la\w,l)$, so the frequency content shrinks to values around $\w=0$, but the spacelike tail due to $V^\fre_a(-\infty,l)=-2\pi\ti{\dV^\fre_a}(0,l)$ (see \eqref{Asp} and \eqref{aV0}) does not change in the limit.\footnote{In the original Staruszkiewicz's article this shrinking of the frequency support is achieved not by scaling, but the introduction of a cutoff function.} Using the semiclassical approximation
for the phase of the wave function of a test particle placed in such field,  Staruszkiewicz found that the incoming plane wave $\exp(-ip\cdot x)$ of a charged particle, with $e$ and $p$ its charge and four-momentum, respectively, acquires in far future a phase shift of the magnitude
\begin{equation}\label{phase}
 \delta(p)=-\frac{e}{2\pi}\int\frac{p\cdot V^\fre(-\infty,l)}{p\cdot l}\,d^2l\,.
\end{equation}
If a wave packet is formed, this shift produces observable effects. It is easy to see their nature. If $f(p)$ is the momentum profile of the initial packet, then the final packet has the profile $e^{i\delta(p)}f(p)$. The addition to the phase has no effect on the distribution of momentum, but under the action of the position operator $-i\p/\p p^a$ causes a shift $\p\delta(p)/\p p^a$.

The same shift is obtained for the trajectory of a classical particle in weak free field $F^\fre_{ab}$, \cite{her17}. For a particle passing through a spacetime point $x_0$, with the four-velocity $v$, which is not affected in the low energy of the field limit, the shift is given by
\begin{equation}\label{deltashift}
 \Delta_a=-\frac{e}{m}\int_\mR F^\fre_{ab}(x_0+v\tau)\tau d\tau\,v^b
 =\frac{e}{2\pi m} \int
 \frac{l_a V^\fre_b(-\infty,l)-l_b V^\fre_a(-\infty,l)}{(v\cdot l)^2}
 \,d^2l\,v^b\,,
\end{equation}
where the first equality follows from an analysis of the equation of motion, and for the second equality we used \eqref{FVint}. The result agrees with the derivative of~$\delta(p)$.
The memory effect for the Dirac field, both classical as quantum, placed in low-energy electromagnetic field was analyzed in \cite{her12}. For the quantum Dirac field one finds that the incoming/outgoing fields $\psi^\inc(\chi)$/$\psi^\out(\chi)$, where $\chi$ is a smearing test bispinor function, are related by the usual similarity $\psi^\out(\chi)=S^*\psi^\inc(\chi)S$, where the scattering operator $S$ has the form
\begin{equation}
 S=\exp\Big[i\int\delta(p)n^\inc(p)d\mu(p)\Big]\,,
\end{equation}
and where $\delta(p)$ is the phase \eqref{phase}, $n^\inc(p)$ is the momentum-density operator of incoming field, with plus/minus sign for electrons/positrons, respectively, and $d\mu(p)$ is the invariant measure on the mass hyperboloid.

More recently, another memory effect has been proposed in electrodynamics \cite{bie13}, which gained much wider popularity. The authors propose to go to `radiation zone' and integrate the electric field over time from infinite past to infinite future. This integral is then supposed to determine a `velocity kick' which a test particle will experience. More precisely, this amounts to the following. Let the spacetime position vectors $x$ be  parametrized by the retarded coordinates, that is $x=ut+Rk$, where $t$ is the unit time axis vector, $k$ is a future null vector such that $t\cdot k=1$, $R\geq0$, and then $u$ is the retarded time. For $R\to\infty$ (radiation zone) the field behaves as in \eqref{nullfut}, and the authors keep only this leading term. Thus, in our language,  the integral postulated by them, apart from the factor $R^{-1}$, is
\begin{equation}\label{int}
\begin{aligned}
 \int_\mR\lim_{R\to\infty}RF_{ab}(ut+Rk)\,du
 &=\int_\mR\big[k_a\dV_b(u,k)-k_b\dV_a(u,k)\big]du\\
 &=k_b V^\out_a(-\infty,k)-k_a V^\out_b(-\infty,k)\,,
\end{aligned}
\end{equation}
where we used \eqref{nullfut}, and then \eqref{VVVV}. The rhs depends on the long-range variable of the outgoing field, that is the difference $V_a(-\infty,k)-V_a(+\infty,k)$, which appears in recent discussions of memory. However, one can raise doubts about actual experimental setting for this calculation \cite{her23_2}. However large $R$ is, it is finite and fixed. Consider, for simplicity, only the case of a free field $F^\fre_{ab}$. For any $R$ (whether large or not), it follows from \eqref{FVint} that
\begin{equation} \label{Int}
 \int_\mR F^\fre_{ab}(ut+Rk)du=0\,,
\end{equation}
as $\dV^\fre_a(\pm\infty,k)=0$. The discrepancy between \eqref{int} and \eqref{Int} results from unallowed, in this case, pulling of $R\to\infty$ limit under the integral \cite{her23_2}. For any finite $R$, to approximate \eqref{int}, the time integration in \eqref{Int} cannot extend over the whole axis. This poses a question of a  definition of the integration interval, and of the way the limit is taken. For $z$ any vector one could try to integrate over a half line:
\begin{equation}\label{timefinite}
 R\int\limits_{\tau_0}^\infty F^\fre_{ab}(\tau t+Rz)d\tau
 =\frac{R}{2\pi}
 \int \big[l_a\dV^\fre_b(\tau_0t\cdot l+Rz\cdot l,l)
 -l_b\dV^\fre_a(\tau_0t\cdot l+Rz\cdot l,l)\big]\frac{d^2l}{t\cdot l}\,.
\end{equation}
For finite $R$ this produces an accidental quantity, which in addition vanishes in the zero energy limit. We try $R\to\infty$ limit in two cases. First, let $z=k\in C_+$, $t\cdot k=1$, and then the limit of the rhs, in analogy to the case of $A^\fre$, takes the form $k_bV^\fre_a(\tau_0,k)-k_aV^\fre_b(\tau_0,k)$.
If $V^\fre(s,l)$ reaches $V^\fre(-\infty,l)$ fast for $s\to-\infty$, then starting integration sufficiently early (but not in minus infinity!) we approximate \eqref{int}. This procedure, experimentally, would have to be rather fine tuned -- integration over retarded time from $\tau_0$ means that with $R\to\infty$ we start integrating over laboratory time $\tau_0+R$ -- tending to infinity.  Moreover, in zero frequency limit $V^{\fre(\la)}(s,l)$ spreads infinitely, so $V(-\infty,k)$ is reached infinitely slowly. Next, consider  $z$ spacelike, then again in analogy to $A^\fre$ the limit of the rhs takes the form
\begin{equation}
 \frac{1}{2\pi}\int[l_bV^\fre_a(-\infty,l)-l_aV^\fre_b(-\infty,l)]\delta(z\cdot l)\frac{d^2l}{t\cdot l}\,.
\end{equation}
If $t\cdot y=0$, then this is obtained by integrating from a fixed laboratory time, and then going far away in a $3$-space direction. This indeed is a functional of $V^\fre(-\infty,l)$, but it also depends on the choice of direction $z$. Anyway, in both cases a finite effect as a function of $V^\fre(-\infty,l)$ does not exist, as it needs infinite scaling.

Finally, we would like to stress that apart from memory effects,
the limits $l_aV_b(\pm\infty,l)-l_aV_a(\pm\infty,l)$ have well defined meaning in general case, and are in principle measurable. As we have seen, for $-\infty$ this quantity is encoded in the spacelike asymptotic behavior of fields \eqref{Fsp}, while for $+\infty$ in the asymptotic trajectories of the outgoing charges.
The important aspect of the adiabatic memory shift discovered earlier is, that for a weak free field it allows to observe its spacelike tail in scattering experiments, without actually going to spatial infinity, as formula \eqref{deltashift} clearly shows---the shift is the result of integrating the field over a timelike line. It is the existence of correlations in the values of $F^\fre_{ab}(x)$---as a solution of the free Maxwell equations---which is responsible for the shift's exclusive dependence on $V(-\infty,l)$. In fact, no limit, whether null or spacelike, is necessary, nor is an infinite scaling.

\subsection{Infrared triangle}

As already alluded at several points above, there is now a renewed activity in the subject of asymptotic properties of electromagnetic fields. This also extends to other massless fields, with stress on gravitational field, in which case hopes are expressed that such analyses may help to address some long standing problems, such as the black hole paradox \cite{haw16}.

In these notes we restrict attention to electrodynamics. The recent activity is organized conceptually by a scheme introduced by A.\ Strominger, named `infrared triangle'. Vertices of this triangle are: asymptotic symmetries/asymptotic charges, soft photon theorems, and memory effects.
At the classical level, these vertices appeared above in our review: charges (\ref{charge}, \ref{chargee}), soft photon relation \eqref{infra}, and memory in Section \ref{mem}. These vertices are expected to have quantum meaning, with charges related to asymptotic symmetries, and we shall refer, and make comments on them in the sequel.  More on the vertices of the triangle, and relations between them, as seen by its author, may be found in review article \cite{str18}.

\subsection{Symplectic structure}\label{sympstr}

Consider the space of free electromagnetic fields with compact initial data on Cauchy surfaces (by finite propagation of fields this property does not depend on the choice of the surface); in this section we omit superscripts `$\fre$', `$\inc$' or `$\out$' indicating free fields in previous sections.  It is well known, and may be easily seen, that this space may be equipped with the symplectic form:
\begin{equation}\label{sympA}
 \sigma(A_1,A_2)=\frac{1}{4\pi}
 \int_\Sigma \big(F_1^{ab}A_{2b}-F_2^{ab}A_{1b}\big)(x)d\sigma_a(x)\,,
\end{equation}
where $\Sigma$ is a Cauchy surface with the dual integration form $d\sigma_a$; if $(\zeta^\mu)$, $\mu=0,1,2,3$ are coordinates parametrizing a neighborhood of $\Sigma$ defined by $\zeta^0=0$, and $\zeta^0$ increases into the future, then $d\sigma_a=\p_a\zeta^0\sqrt{|g_\zeta|}d^3\zeta$, where $g_\zeta$ is the determinant of the metric tensor in coordinates $\zeta^\mu$. By Stokes' theorem (and free Maxwell equations) this form is gauge-independent, and by Gauss' theorem it is independent of the choice of the Cauchy surface $\Sigma$. If $A_{1,2}$ are Lorenz potentials, then by the use of Stokes' theorem one can replace the integrand by
$A_{2b}\,\p^a\!A_1^b-A_{1b}\,\p^a\!A_2^b$, and then one finds that $\sigma(A_1,A_2)=\{V_1,V_2\}$, where
\begin{equation}\label{sympV}
 \{V_1,V_2\}=\frac{1}{4\pi}
 \int\big(\dV_1\cdot V_2-\dV_2\cdot V_1\big)(s,l)\,dsd^2l\,;
\end{equation}
this may be obtained either by the explicit use of representation  \eqref{AVint} (and integration as in Appendix C of \cite{her95}), or by shifting the Cauchy surface to the future null infinity. The latter interpretation transfers the discussion of fields to the future null infinity, and this has been done by various authors before, notably Bramson \cite{bra77} and Ashtekar \cite{ash81,ash86}. For initial data, as assumed,  of compact support on $\Sigma$, the asymptotes $V$ are also of compact support (by the lightlike propagation of solutions of the wave equation). However, the form $\{.,.\}$ has the obvious, straightforward  extension to all free fields from the class of Section \ref{nullspace} (the integrand in \eqref{sympV} is still absolutely integrable). Because of the crucial role of a symplectic structure for quantization, this is of key importance for the possibility of a quantum theory for these fields. In contrast, functions of the decay \eqref{gc}, characteristic for fields with even spacelike decay \eqref{FFsp} are not admitted.\footnote{Anticipating what comes in Section \ref{star}, we mention that it is possible to define a symplectic structure on the space of all fields of Coulomb-rate, electric type decay, but at the price of neglecting their spacetime behavior, and taking into account only their spacelike tails.}  We note that because of orthogonality of $V$'s to $l$, the symplectic form is gauge invariant.

There is still another way to represent the symplectic structure \eqref{sympA}. Suppose that potentials $A_i$, $i=1,2$, are represented as radiation potentials of some conserved test currents $J_i$, with asymptotic behavior as for $j$ in Section \ref{nullspace}
\begin{equation}
 A_{ia}(x)=4\pi\int D(x-y)J_{ia}(y)dy
 =-\frac{1}{2\pi}\int \dV^{J_i}_a(x\cdot l,l)\,d^2l\,,
\end{equation}
where $\dV^{J_i}_a(s,l)$ are formed as in \eqref{Vj}. Then
 $\sigma(A_1,A_2)=\{J_1,J_2\}_c$ (subscript $c$ for `current') where
\begin{equation}\label{sympJ}
 \{J_1,J_2\}_c=\tfrac{1}{2}\int [J_1\cdot A_2-J_2\cdot A_1](x)dx\,.
\end{equation}
If $J_i$ are local currents, then this reduces to the standard form
\begin{equation}\label{sympJloc}
 \{J_1,J_2\}_c=4\pi\int J_1^a(x)D(x-y)J_{2a}(y)\,dxdy\,,
\end{equation}
but for more general currents the latter representation is not absolutely integrable.

We end this section by remarking that under the transformation \eqref{VVinf}
applied to $A_1$, $A_2$, the symplectic structure \eqref{sympV} is
transformed by
\begin{equation}
 \{V_1, V_2\}\mapsto  \{V_1, V_2\}
 -\frac{1}{4\pi}\int \big[V_1(-\infty,l)\cdot V_2^+(l)
 -V_2(-\infty,l)\cdot V_1^+(l)\big]d^2l\,,
\end{equation}
which again shows that this is not a gauge transformation, at least within the structure reviewed here, on which quantization to be discussed in the following section is based. Possibility of another interpretation of the above transformation, within a broader scheme---that of BV-BFV formalism---has been recently presented in \cite{rej21}.

\section{Quantum theory beyond locality}\label{quant}

\subsection{Vacuum representation of local electromagnetic fields}\label{vac}

The standard local quantization of the free electromagnetic field may be formulated as a mapping from the space of smooth, conserved, compactly supported currents $J$ to operators $A(J)$ in a Hilbert space, satisfying commutation relations
\begin{equation}\label{comloc}
 [A(J_1),A(J_2)]=i\{J_1,J_2\}_c\,,
\end{equation}
with $\{J_1,J_2\}_c$ as in \eqref{sympJloc}. As $J_a$ are conserved, the elements $A(J)$ are gauge invariant, so they represent the smeared electromagnetic field. We denote
\begin{equation}\label{JV}
 V_{ia}(s,l)=V_a^{J_i}(s,l)=\int \delta(s-x\cdot l)J_{ia}(x)dx\,,\quad i=1,2\,,
\end{equation}
which functions are homogeneous of degree $-1$, orthogonal to $l$ (as the charge $q=0$ for compactly supported current).
For smearing currents of compact support these functions are also of compact support. We write $A(J)=\{V, V^\qu\}$, and then this quantity symbolizes a quantum (operator) asymptote $V^\qu(s,l)$ smeared according to \eqref{sympV} with a test profile $V(s,l)$. In terms of these elements \eqref{comloc} takes the form
\begin{equation}\label{comreg}
 \big[\{V_1, V^\qu\},\{V_2, V^\qu\}\big]=i\{V_1, V_2\}= -\int \bigg\{\int_\mR \ov{\ti{\dV_1}}\cdot\ti{\dV_2}(\w,l)\frac{d\w}{\w}\bigg\}d^2l\,,
\end{equation}
where $\ti{\dV}(\w,l)$ are defined as in \eqref{aV}. The usual vacuum Fock representation of the free electromagnetic field may be then formulated as follows. On the space of functions $\ti{\dV}(\w,l)$ one defines the scalar product
\begin{equation}\label{scpro}
 (\ti{\dV_1},\ti{\dV_2})=-\int \bigg\{\int_0^\infty \ov{\ti{\dV_1}}\cdot\ti{\dV_2}(\w,l)\frac{d\w}{\w}\bigg\}d^2l\,;
\end{equation}
as $V_{1,2}(s,l)$ are of compact support, so $\ti{\dV_{1,2}}(\w,l)\propto \w$ for $\w\to0$, and the product is well defined. The completion of this space with respect to this product defines a Hilbert space $\Hc_1$. One forms now the symmetrized Fock space $\Hc_F$ based on this `one-excitation' space, and defines
\begin{equation}
 \{V, V^\qu\}=a(\ti{\dV})+a^*(\ti{\dV})\,,
\end{equation}
where $a(\ti{\dV})$ and $a^*(\ti{\dV})$ are annihilation and creation operators in $\Hc_F$, respectively, satisfying relations
\begin{equation}
 \big[a(\ti{\dV_1}), a^*(\ti{\dV_2})\big]=(\ti{\dV_1},\ti{\dV_2})\,,\qquad
 a(\ti{\dV_1})\W=0\,,
\end{equation}
where $\W$ is the Fock vacuum. As
$
 (\ti{\dV_1},\ti{\dV_2})-(\ti{\dV_2},\ti{\dV_1})= i\{V_1, V_2\}
$,
relations \eqref{comreg} follow. One also shows that there is a unitary representation of the Poincar\'e group in action in $\Hc_F$, which transforms these elements covariantly (for the precise meaning of that see the next subsection). In particular, the joint spectrum of generators of translations---the four-momentum operators---covers the solid future light cone, as expected on physical grounds. Note that for $V(-\infty,l)=-2\pi\ti{\dV}(0,l)\neq0$ the product \eqref{scpro} diverges in $\w\searrow0$, so such fields---infrared singular---are not admitted.

We have seen in Section \ref{nullspace} that in order to describe classical scattering one has to admit $V$'s with $V(-\infty,l)\neq0$. On the other hand, the above standard algebra does not encompass quantum analogs of such fields. To better understand this, consider relation
\begin{equation}\label{translV}
 e^{i\{V_1,V^\qu\}}\{V,V^\qu\}e^{-i\{V_1,V^\qu\}}=\{V, V^\qu\}+\{V, V_1\}\,,
\end{equation}
which follows from relations \eqref{comreg}. This has the following interpretation: the similarity transformation on the lhs adds to the quantum field, with quantum asymptote $V^\qu$, another field with asymptote $V_1$. In particular, take the quantum mean value of both sides in the vacuum state, $(\W,.\,\W)$. The vacuum mean value of $\{V, V^\qu\}$ vanishes, and one obtains $\{V, V_1\}$, $V_1$ smeared with the test profile $V$. However, as $V_1(s,l)$ is of compact support, no scattered field can be obtained.

Let us now consider the rhs of \eqref{translV}, but with $V_1$ not of compact support, but instead any function from the extended symplectic space of the last section. The addition $\{V, V_1\}$ is still well-defined, but it cannot be obtained as on the lhs of this formula. Nevertheless, the elements defined by
$\{V, V^{\qu\prime}\}=\{V, V^\qu\}+\{V, V_1\}$, with fixed $V_1$, again satisfy commutation relations \eqref{comreg}, so they form a representation of the free electromagnetic field, in which the spacelike asymptote of the electromagnetic field is given by a classical function \eqref{Fsp} determined by $V_1(-\infty,l)$. These coherent representations, as they are called, are inequivalent for different limit functions $V_1(-\infty,l)$, in particular they are not equivalent to the vacuum representation for $V_1(-\infty,l)\neq0$.

\subsection{Extended algebra of free electromagnetic field}\label{extended}

The inability of the local vacuum representation to accommodate radiation fields produced in scattering (as described at the end of the last subsection),  formulated in various ways, is a very old problem, which has been addressed early and by many authors. The common strategy of solutions proposed within the limits of local theory, in various guises, is the addition of classical currents, depending on particle momenta, with the aim of correcting the deficit of the long range tails. Here is, among others, the place of the Faddeev-Kulish approach, also in its newest reformulation \cite{duc19}, described in \cite{duc23}.

In all treatments mentioned above, the algebra of free electromagnetic fields remains the same, that given by local elements \eqref{comloc}.
One can ask whether indeed the long range variables of electromagnetic fields have only the classical auxiliary character, as described above. Again, it seems that Staruszkiewicz was the first author to pose this question, after observability of these variables has been indicated in his memory effect. However, his proposition is based on the extracted long range variables per se, and it does not fit into the present setting, so we postpone its description to Section~\ref{star}.

Here we briefly describe an extended algebra based on the symplectic forms \eqref{sympV} or \eqref{sympJ}. We note that these forms have natural extensions to infrared singular fields. These extensions have been used in \cite{her98} to construct a wider algebra of fields, including those infrared singular, and a class of its representations (see also \cite{her17,her23_1} and references therein). The extension of the form \eqref{sympV} has been earlier considered  by Ashtekar \cite{ash86}, but only for purposes of a Faddeev-Kulish type construction, and not for extending the algebra.

We describe the extension for the case of the free `out' field (but we omit the superscript), the `in' case is similar. For the outgoing field it is natural to use the future asymptotic test functions $V_a(s,l)$. The main point in the extension of the symplectic form \eqref{sympV} is that one admits test functions $V_a(s,l)$ with $V_a(-\infty,l)\neq0$, while it remains true that $V_a(+\infty,l)=0$. The functions are assumed smooth, and $\dV_a(s,l)$, together with all their derivatives by $L_{bc}$, decay at least as $|s|^{-1-\vep}$ in infinity. Moreover, $L_{[ab}V_{c]}(-\infty,l)=0$ and $l\cdot V(s,l)=0$, which together imply that there exists a scalar function $\Phi(l)$, determined up to the addition of a constant, such that $l_aV_b(-\infty,l)-l_bV_a(-\infty,l)=L_{ab}\Phi(l)$. The space of such functions with the form \eqref{sympV} is a nondegenerate symplectic space. We are looking for a representation of elements $\{V, V^\qu\}$ as operators in some Hilbert space, which satisfy \eqref{comreg} for this extended symplectic space. However, as there are no such representations by bounded operators (as even in the simplest case of one canonical pair $[X,P]=i\id$), it is mathematically more manageable to start with the usual Weyl translation of such canonical problem to an abstract $C^*$-algebraic problem.

We assume existence of elements $W(V)$ which satisfy abstract algebraic relations
\begin{equation}\label{asalg}
 W(V_1)W(V_2)=e^{-\frac{i}{2}\{V_1,V_2\}}W(V_1+V_2)\,,\quad W(V)^*=W(-V)\,.
\end{equation}
According to general results (see e.g.\ \cite{bra96}, Thm. 5.2.8) there exist (as for each nondegenerate symplectic form $\{.,.\}$ on a vector space) a unique, simple, $C^*$-algebra, called Weyl algebra, generated by such elements; we shall denote our algebra $\Ac$. Therefore, there exist representations $\pi:\Ac\mapsto \Bc(\Hc)$ from the algebra to the space of bounded operators on some Hilbert space $\Hc$, which map generators $W(V)$ of the algebra to unitary operators $\pi(W(V))$, denoted in the following by $W_\pi(V)$ for simplicity, satisfying these relations. For a given $C^*$-algebra there usually exist many inequivalent representations, and $\pi$ denotes any of them. All representations are faithful, due to simplicity of $\Ac$, and we demand in addition irreducibility. Not all such representations have physical significance, and one has to  apply some selection criteria. In the case of Weyl algebra one of the most important is regularity, in the sense that for each $V$ the unitary one-parameter group $\mR\ni\tau\to W_\pi(\tau V)$ is strongly continuous, and then $W_\pi(V)=\exp[-i\{V, V^\qu\}_\pi]$, with self-adjoint operators $\{V,V^\qu\}_\pi$ defined by this relation. Then for each pair $V_1, V_2$ there exists a dense subspace of the representation space on which relations \eqref{comreg} are satisfied (a consequence of the Stone--von Neumann uniqueness theorem, see \cite{bra96}). As the symplectic form \eqref{sympV} is gauge invariant, as described in Section \ref{sympstr}, the operators $W_\pi(V)$ are gauge invariant and represent exponentiated electromagnetic fields.

Further possible selection criteria refer to the action of symmetries on the algebra. For $(a,\La)$ symbolizing an element of the Poincar\'e group (with $a$ a translation vector, and $\La$ a Lorentz transformation), the mapping $(a,\La)\mapsto \al_{a,\La}$, with $\al_{a,\La}$ the ${}^*$-automorphism of the algebra given by
\begin{equation}
 \al_{a,\La}(W(V))=W(T_{a,\La}V)\,,\qquad
 [T_{a,\La}V]^a(s,l)=\La^a{}_b V^b(s-a\cdot l,\La^{-1}l)\,,
\end{equation}
defines a representation of the Poincar\'e group by ${}^*$-automorphisms of the Weyl algebra. One says that such automorphisms are implementable in the representation $\pi$, if there is a unitary representation $U(a,\La)$ of the Poincar\'e group, in action in the representation space $\Hc$, such that
\begin{equation}
 U(a,\La)W_\pi(V)U(a,\La)^*=W_\pi(T_{a,\La}V)\,.
\end{equation}
If this is the case, the representation is called Poincar\'e-covariant (and similarly for restrictions of the Poincar\'e group to the Lorentz group or the group of translations). And finally, if the representation is translation-covariant, strongly continuous in $a$, then the generators $P_a$ of translations, \mbox{$U(a)=\exp[ia\cdot P]$}, have the interpretation of four-momentum operators. Relativistic condition for the positivity of energy takes then the form of the selection criterion demanding that the joint spectrum of $P_a$ is contained in the closed, solid future light cone, and the representation is then called positive energy representation.

We note that if $V$'s are restricted to those compactly supported, then the algebra \eqref{asalg} reduces to the algebra of local fields described in Section \ref{vac}, and the vacuum Fock representation reviewed there satisfies all selection criteria described above: it is an irreducible, regular, Poincar\'e-covariant, positive energy representation. However, by general results (see \cite{bra96}, Thm. 5.2.9), this restriction produces a subalgebra strictly smaller than the full algebra. Moreover, the vacuum representation may be extended to a larger subalgebra, namely that formed by all $W(V)$ with $V(-\infty,l)=0$ (the scalar product \eqref{scpro} remains well defined), but makes no sense for infrared singular profiles $V$.

There is one more general point one should mention. While $\pi(\Ac)$, for any representation $\pi$, is ${}^*$-isomorphic to $\Ac$ (all $\pi$ are faithful), and as a subspace of $\Bc(\Hc)$ is closed in the norm topology, of physical interest in the given representation are also elements which may be reached from $\pi(\Ac)$ by strong (equivalently: weak) topology, which produces a von Neumann algebra $\pi(\Ac)''$, bicommutant of $\pi(\Ac)$, as well as self-adjoint operators with spectral projections in $\pi(\Ac)''$. For a regular representation, in the latter category are, in particular, generators $\{V, V^\qu\}_\pi$ of $W_\pi(\tau V)$.

\subsection{Representations}\label{representations}

Further development of the theory may be here only roughly outlined, and we refer the reader to original articles for details.
A class of translationally covariant, positive energy representations of the extended algebra has been constructed in \cite{her98}; we shall now write $\pi=r$ to denote these representations. The idea of the construction is the following. First, restrict the algebra to the standard infrared regular fields, and consider its coherent representations, as briefly characterised at the end of Section~\ref{vac}, with all possible functions $V_1(-\infty,l)$ characterizing them. Then choose a Gaussian measure, with appropriate regularity properties, on the space of these functions, and form the direct integral Hilbert space $\Hc_r$ of these representations. On the integrated Hilbert space the full algebra \eqref{asalg} has a natural, regular, irreducible and translation-covariant representation, extending the infrared regular subalgebra. The energy-momentum spectrum covers the whole (solid) closed future light cone, but there is no vacuum state. Instead, there is a distinguished class of  vector states $\W_h\in\Hc$ with arbitrarily low residual energy content (their index functions $h$ determine, among others, this content). Representations thus formed may be chosen as covariant with respect to $3$-space rotations, but it does not seem possible to have full Lorentz covariance. However, the long range degrees of freedom have fully quantum character.

The direct integral construction of the representations ensures the covariance properties and the energy spectrum structure described above. However, of both fundamental and technical importance is the fact that the states $(\W_h,.\,\W_h)$ are Fock states, so these representations are unitarily equivalent to certain Fock representations (not the vacuum Fock representation of Section \ref{vac}). As a consequence, these representations have further regularity properties not following from the strong continuity of one-parameter groups $W_r(\tau V)$ alone; among them is the existence of a dense subspace $\Dc\subset\Hc$, which is a domain of essential self-adjointness of all elements $\{V,V^\qu\}_r$, invariant with respect to them, and on which the commutation relations \eqref{comreg} are satisfied. Another important consequence is the possibility of the construction, in the von Neumann algebra $r(\Ac)''$, of the unitary transformations generated by quantum representations of charges of Section \ref{matchar}, see below.

The ideas on the infrared structure described in Introduction, and their concrete realisation in the form of the algebra \eqref{asalg}, have been recently formulated anew in Ref.\ \cite{pra22}. The article puts as one of its main objectives to find a Poincar\'e covariant representation of the algebra \eqref{asalg}, and in this respect its conclusion is negative---no such representation has been constructed. From the point of view of our earlier constructions this can hardly be a surprise. The (homogeneous) Lorentz covariance may be achieved for a theory of the long range electromagnetic tails per se, as in the Staruszkiewicz model (see the next section), but then the problem reappears in attempts to extend such model to the whole spacetime. Thus, the obstacle to a Poincar\'e covariant representation of all fields, including the infrared singular ones, seems to rest in the demand to reconcile the (homogeneous) Lorentz covariance with the positive energy representation of translations. The lack of the full Lorentz covariance is a known problem in local theory, as described in Introduction, and it seems to reappear here in a different guise.

\subsection{Asymptotic extended Dirac-Maxwell algebra and further developments}\label{mdext}

Consider now a system of interacting Dirac particles and electromagnetic field. Based on the assumptions that the outgoing Dirac particles tend to free particles carrying their own Coulomb field (identical with advanced field), the following ${}^*$-algebra has been proposed in \cite{her98} as an asymptotic algebra of this system in future;\footnote{One can also turn it into a $C^*$-algebra by an appropriate choice of a norm, but we shall not need this here.} similar construction with future changed to past gives then the asymptotic algebra in past. Algebra is generated by elements $W(V)$,  with test functions and algebraic relations as in \eqref{asalg}, and  elements $\psi(g)$ symbolizing Dirac field, where $g$ is a bispinor Schwartz test function on the future unit hyperboloid
$H_1=\{v\in M|\,v^2=1, v^0>0\}$; in this formulation $g(p/m)$ is the restriction to the mass hyperboloid $p^2=m^2$, $p^0>0$, of the Fourier transform of a spacetime smearing function for $\psi(x)$.  The Dirac field anticommutation relations take the form
\begin{equation}\label{dirac}
 \big[\psi(g_1),\psi(g_2)\big]_+=0\,,\quad
 \big[\psi(g_1),\psi(g_2)^*\big]_+=(g_1,g_2)=\int \ov{g_1}(v)\gamma\cdot v g_2(v)d\mu(v)\,,
\end{equation}
where bar denotes the Dirac conjugation, $\gamma^a$ are the Dirac matrices, and $d\mu(v)=d^3\vv/v^0$ is the invariant measure on $H_1$. The elements $W(V)$ now symbolize the total field, and for its test function $V$ we denote
\begin{equation}\label{ssig}
 S_\Phi(v)=e^{\txt i\Sigma_\Phi(v)}\,,\quad
 \Sigma_\Phi(v)
 =\frac{1}{4\pi}\int V^e(v,l)\cdot V(-\infty,l)\,d^2l
 =\frac{e}{4\pi}\int\frac{\Phi(l)}{(v\cdot l)^2}\,d^2l \,,
\end{equation}
where $V^e(v,l)=ev/(v\cdot l)$ is the advanced potential asymptote of the advanced field of a free particle with charge $e$ and four-velocity $v$, and $\Phi(l)$ and $V(-\infty,l)$ are related as in (\ref{fsp}--\ref{Vfi}). The following further assumed relations
\begin{equation}\label{asalgD}
 W(V)\psi(g)=\psi(S_\Phi g)W(V)
\end{equation}
take care of the fact that creation/anni\-hil\-ation of the charged particle adds/subtracts its Coulomb field to the total field (this will be better seen at the level of representations, below).
Change of gauge $V(-\infty,l)\mapsto V(-\infty,l)+l\al(l)$ causes a shift of $\Sigma_\Phi(v)$ by a constant, which results in a constant phase shift in $S_\Phi g$ \eqref{asalgD}.

We construct representations of relations \eqref{asalg}, \eqref{dirac} and \eqref{asalgD} as follows.  We denote by
\begin{equation}
 \psi_D(g)=(g,g^\qu)_D=\int\ov{g}(v)\gamma\cdot v\,g^\qu(v)d\mu(v)
\end{equation}
the standard vacuum Fock representation of the Dirac field in the Hilbert space $\Hc_D$, with $g^\qu(p/m)$ symbolizing Fourier transformed quantum field. We further form
\begin{equation}
 n^\qu(v)=\,:\ov{g^\qu}(v)\gamma\cdot v g^\qu(v):\,,\qquad
 V^\qu(+\infty,l)=\int V^e(v,l)n^\qu(v)d\mu(v)\,,
\end{equation}
with normal product in the first formula; $n^\qu(v)$ has the interpretation of the $4$-velocity density of particles, with plus/minus sign for electrons/positrons, respectively, and $V^\qu(+\infty,l)$ is the quantum asymptote of the advanced potential. Let $W_r(V)$ be a representation on $\Hc_r$, as briefly characterized in Section \ref{representations}. Then the following elements
\begin{equation}
 \psi_\pi(g)=(g, g^\qu)_D\otimes\1\,,\quad
 W_\pi(V)=\exp\Big\{\frac{i}{4\pi}\int V(-\infty,l)\cdot V^\qu(+\infty,l)\,d^2l\Big\}\otimes W_r(V)
\end{equation}
satisfy the assumed asymptotic algebraic relations. If we now write
\begin{equation}
 W_\pi(V)=e^{-i\{V, V^\qu(+\infty,.)\}}\otimes e^{-i\{V,V^{\qu\out}\}_r}\,,
\end{equation}
with the form of the first factor easily obtained, we see that
 $V^\qu(s,l)=V^\qu(+\infty,l)\otimes\1+\1\otimes V^{\qu\out}(s,l)$ describes the
quantum asymptote of the total field.

Once we have the asymptotic algebra and its representations, we can ask whether the quantum analogs of charges $Q(V^+)$, or better their exponentiations $U(V^+)=\exp[-iQ(V^+)]$, may be constructed. On the classical level, it is clear that $U(V^+)$ cannot be expressed in terms of $\{V_1,V\}$ for some fixed $V_1(s,l)$, and some limiting must be involved. For instance, if we assume that $V_1(-\infty,l)=V^+(l)$, and denote $V_{1\tau}(s,l)=V_1(s-\tau t\cdot l,l)$, then $\dsp \lim_{\tau\to-\infty}|U(V^+)-W(V_1)|=0$. In the quantum case, any similar limit does not exist in the norm limit sense, as for
$V_1\neq V_2$ one has $\|W_\pi(V_1)-W_\pi(V_2)\|=2$, as is the case in any Weyl algebra. Therefore, the algebraic structure alone is not sufficient, and one has to use its concrete Hilbert space representation, and consider weaker limiting. As it turns out, in the representations defined above the strong limit of $W_\pi(V_{1\tau})$ does not exist either. However, there does exist limiting procedure, in strong topology, which produces in the von Neumann algebra operators $U(V^+)$ with clear interpretation of analogs of classical expressions, for details we refer the reader to \cite{her17,her98}. These operators have the structure $U^{\mathrm{Coul}}(V^+)U^{\mathrm{free}}(V^+)$, with the first factor being identity in $\Hc_r$, and the second in $\Hc_D$; in the language used recently these factors are exponentiations of  `hard' and `soft photon' charges, respectively. These operators induce transformation of the basic variables of the representation given by
\begin{align}
 U(V^+)W_\pi(V)U(V^+)^*
 &=\exp\Big\{\frac{i}{4\pi}\int V^+(l)\cdot
 V(-\infty,l)\,d^2l\Big\}\, W_\pi(V)\,,\label{lrW}\\
 U(V^+)\psi_\pi(g)U(V^+)^*&=\psi_\pi(S_{\vep^+}g)\,,\label{lrP}
\end{align}
where $\vep^+$ and $V^+$ are related as in (\ref{fsp}--\ref{Vfi}).
These relations show that the transformation is not a pure gauge: it changes the phase of the Dirac field, for which the `hard' part is responsible, but it also changes the electromagnetic field by adding an advanced charge-free field, for which `soft' part is responsible. This is in agreement with the classical analysis of Section \ref{gauge}.

The above construction of the asymptotic algebra is physically motivated, and may be regarded as a step in the direction outlined in Introduction. Further investigations plan should aim at a construction of interacting theory, at least in some perturbative steps, in which these asymptotic structures will have a natural place. Such work is in progress, and preliminary steps in this direction have been proposed recently. First, on the classical level, it was shown that the Dirac field, scattered in external electromagnetic field, approaches its free versions in far past and future, without any Dollard corrections, if an appropriate gauge is chosen \cite{her21}. The constitutive property of such gauge is that $x\cdot A(x)$ vanishes sufficiently fast in past and future. Next, motivated by this result, a special gauge with this property---the almost radial gauge---has been constructed\footnote{The construction may be interpreted as an integration of the electromagnetic field tensor along radial lines going out from points in a neighborhood of $x=0$ (over which, subsequently, some average is taken). This may resemble the idea of string-localized fields of \cite{mun06}, but the similarity is rather superficial; we discuss it in \cite{her22_2}.} both classically, and in the representation $\pi=r$ of the algebra \eqref{asalg}, \cite{her22_2}. Finally, this algebra and the almost radial gauge have been coupled with the Dirac field, for the construction of the lowest order perturbation calculus in QED, \cite{her23_1}. A~few results of these investigations, not available in the standard local theory, could be mentioned. The scheme avoids indefinite metric, and allows the construction of the Dirac field as an operator in the Hilbert space up to the first order. This field tends strongly, in far past, on a dense subspace, to the free `in' field. This is not in contradiction with the absence of a discrete mass hyperboloid in the spectrum, as here there is no vacuum with discrete vertex of the energy-momentum light cone in the spectrum. Spacelike asymptotic electromagnetic field may be obtained as a quantum field, and the fluctuations remain bounded in the spacelike limit. Thus, the discrete mass in the energy-momentum transfer of the Dirac field is not achieved by masking the changes of the field in spacelike infinity by fluctuations, as is the case in infravacuum representations.\footnote{For more information on infravacuum representations see \cite{duc23}.} Causal asymptotic limits of fields seem to justify the asymptotic structure used above in this section. There exist reasons to expect that higher perturbation orders
will not destroy the general picture, but their construction demands a solution of the UV problem, which remains a subject for further research.
Within such prospective theory, one should be able to construct asymptotic charges both in the language of incoming and outgoing fields. The equivalence of these two constructions would then supply precise formulation for problems considered on heuristic level in infrared triangle's vertices, see below.

\subsection{Quantum infrared triangle}

It is recently often claimed, as reviewed in \cite{str18}, that quantum charges $Q(V^+)$ generate large gauge transformations, and this, in fact, is one of the vertices of Strominger's infrared triangle. These claims are based on formal treatment of the algebraic relations. This, as we have seen, is not sufficient for mathematical precision---one needs to consider particular representations. In our opinion, as expressed in Section \ref{mdext}, these transformations, in addition to a phase change of the Dirac particle, induce a physical effect in the electromagnetic field.

Another two vertices of Strominger's triangle are also related to the conserved `charges'. Suppose that a full interacting theory may be constructed, with the algebra described above as the outgoing algebra, and an analogous incoming algebra. Then based on classical analysis one should expect that the charges formed in the `out' algebra should agree with those formed in the `in' algebra. This, in Strominger's interpretation (but in our notation of asymptotes), should take the form
 \begin{equation}\langle\out|V(-\infty,l)S|\inc\rangle
 =\langle\out|SV'(+\infty,l)|\inc\rangle\,,
\end{equation}
where $S$ is the scattering operator, the incoming and outgoing states involve charged particles as assumed before relation \eqref{infra} (see \cite{pas17,str18}), and $V(-\infty,l)$ and $V'(+\infty,l)$, are quantum variables, which after smearing with $V^+(l)$ give the charges in terms of outgoing or incoming variables, respectively. If for $V(-\infty,l)$ and $V'(+\infty,l)$ one substitutes some formal `quantization' of the left and right hand sides of \eqref{infra}, respectively, and also takes into account relations \eqref{aVoutin}, then one claims that
\begin{equation}\label{weinb}
 \langle\out|\int\dV(s,l)ds\,S|\inc\rangle=
 -2\pi \lim_{\w\searrow0}\w \langle\out| a^\out(\w l)S|\inc\rangle
 =\Big[\sum_{i=1}^{n}q_i\frac{v_i}{v_i\cdot l}
 -\sum_{i=1}^{n'}q'_i\frac{v'_i}{v'_i\cdot l}\Big]
 \langle\out|S|\inc\rangle\,,
\end{equation}
for the case that $\dsp\lim_{\w\searrow0}\w a^\inc(\w l)|\inc\rangle=0$ (no `soft photons' in the initial state). The integral on the lhs is related to what in that literature is called `memory'.  Similarly, taking the limit $\w\nearrow0$ one would obtain an analogous relation with the creation operator acting on $|\inc\rangle$, and outgoing `memory' replaced by that of $V'$ (for the case of no soft photons in the outgoing state). These relations, in interpretation of \cite{str18}, form a link between two remaining (in addition to LGT) vertices of infrared triangle: quantum memory on the lhs of \eqref{weinb}, and Weinberg's soft photon theorem, the second equality in this relation.

Again, relations of this type have only a heuristic value, as no consistent theory of scattering, with well defined scattering operator $S$, or creation/annihilation operators, stands behind these formal manipulations.
However, these ideas recently arise inspiration also in the mathematical community, see e.g.~\cite{dyb19}.

\section{Staruszkiewicz's theory of quantum electromagnetic fields in spacelike infinity}\label{star}

Staruszkiewicz formulated his theory in Ref.\ \cite{sta89} in a set of axioms on the vacuum and quantum variables of the theory. Later it was shown that the mathematics of the theory is best formulated as a $C^*$-Weyl-algebra model, and its particular representation \cite{her05,her22_1}. We present the theory in terms used in Ref.\ \cite{her22_1}.

The starting point for the theory is a classical scalar field $S(x)$, satisfying the wave equation $\Box S(x)=0$, and homogeneous of degree $0$, which Staruszkiewicz interprets as a phase function. General solution satisfying these assumptions has the form
\begin{equation}\label{corresp_S}
 S(x)=-\frac{e}{4\pi}\int \bigg\{c(l)\sgn(x\cdot l)
 +\p^2 D(l)\log\Big(\frac{|x\cdot l|}{v\cdot l}\Big)\bigg\}d^2l+S_v\,,\quad
 S_v=\frac{e}{4\pi}\int\frac{D(l)}{(v\cdot l)^2}\,d^2l\,,
\end{equation}
where $c(l)$ and $D(l)$ are functions homogeneous of degree $-2$ and $0$, respectively, $e$ is a constant which will be interpreted as the elementary charge, and the solution is independent of the choice of an arbitrary four-velocity $v$. If  for $x^2<0$ we define the potential $A_a(x)=-x_aS(x)/(ex^2)$, then the corresponding field $F_{ab}(x)$ gives the general solution of free Maxwell's equations in $x^2<0$, homogeneous of degree $-2$, of electrical type
\begin{align}
 F_{ab}(x)&=\frac{1}{8\pi x^2}\int
 \frac{l_ax_b-l_bx_a}{x\cdot l-i0}
 \Big[\p^2 D(l)-i\tfrac{2}{\pi}c(l)\Big]\,d^2l+\mathrm{c.c.}\label{Sch}\\
 &=\frac{1}{8\pi}\int\frac{L_{ab}\big[D(l)-i\tfrac{2}{\pi}F(l)\big]}
 {(x\cdot l-i0)^2}\,d^2l+\mathrm{c.c.}\,,\label{Sfree}
\end{align}
where the second representation holds in case that $c(l)=\p^2F(l)$ for a homogeneous function $F(l)$; in this case formula \eqref{Sfree} is already known to us from \eqref{FFsp}. The general term $c(l)$ has for each four-velocity $v$ the representation
\begin{equation}\label{cQ}
 c(l)=\frac{Q(c)}{(v\cdot l)^2}+\p^2F_v(l)\,,\quad Q(c)=\frac{1}{4\pi}\int c(l)\,d^2l\,,
\end{equation}
where $Q(c)$ is the charge of the field \eqref{Sch}, as calculated by Gauss' law, and $F_v(l)$ is a $v$-dependent homogeneous function of $l$. One can show that the term $Q(c)/(v\cdot l)^2$, when used in \eqref{Sch}, reproduces field \eqref{Fsp} for $V_b(-\infty,l)=Q(c)v_b/v\cdot l$, which indeed supplements \eqref{Sfree} to a general field homogeneous of degree $-2$.

The function $S(x)$ may be restricted to the de~Sitter space $x^2=-1$, and then the wave equation in $x^2<0$ and homogeneity imply the wave equation inside the de~Sitter space. Local, in the de~Sitter space, quantization of this field is achieved by putting
\begin{equation}
 \Big[\frac{1}{4\pi}\int\hat{c}(l)D(l)\,d^2l,\frac{1}{4\pi}\int\hat{D}(l')c(l')\,d^2l'\Big]
 =\frac{i}{4\pi}\int D(l)c(l)\,d^2l\equiv i\langle D,c\rangle\,,
\end{equation}
where the `hatted' functions are quantum variables, and those not hatted are test functions, all with homogeneity indicated by $c$ and $D$ as before. It follows from this relation, in particular, that $[\hat{Q},\hat{S}_v]=ie$. This confirms the interpretation of $\hat{S}_v$ as a phase operator, but it also implies that it should be used only in the exponentiated form, and then the commutation relation takes the form
 $\hat{Q}\exp(-i\hat{S}_v)=\exp(-i\hat{S}_v)(\hat{Q}+e\id)$. These heuristic considerations lead to the following algebraic formulation of the quantum theory.

The algebra consists of elements $W(D,c)$, where $D$ and $c$ are test fields as defined above, but with the restriction that $Q(c)=ne$, $n\in\mZ$, see \eqref{cQ}. These elements are assumed to satisfy
\begin{equation}\label{alg_weyl}
\begin{gathered}
 W(D_1,c_1)W(D_2, c_2)
 =\exp\big[\tfrac{i}{2}\sigma(D_1, c_1; D_2, c_2)\big]
 W(D_1+D_2,c_1+c_2)\,,\\[1.5ex]
 W(D, c)^*=W(-D, -c)\,,\qquad W(0,0)=\1\,,
 \end{gathered}
\end{equation}
where $\sigma(D_1,c_1;D_2,c_2)
=\langle D_1,c_2\rangle-\langle D_2,c_1\rangle$. These relations define the algebra as a Weyl algebra over a symplectic Abelian group, and theorems \cite{man73} guarantee its unique existence as a $C^*$-algebra. The element $\exp(-i\hat{S}_v)$ of our earlier discussion corresponds, in this precise formulation, to the element $W(0,c_v)$, with $c_v(l)=e/(v\cdot l)^2$. Thus,  $W(0,c_v)$ creates a charged field, with charge $e$, rotationally symmetric in the frame with the time axis along $v$ (a Coulomb field).

It is known that a scalar field on the de Sitter space (not a phase field) has a one-parameter family of disjoint irreducible representations covariant with respect to the homogeneous Lorentz group in the ambient space, each with an invariant vacuum state \cite{che68}. As it turns out, the same is true for the above algebra \cite{cas11,her05}. Staruszkiewicz picks out a particular representation $\pi$ as follows. Within the theory one can form the quantum analog of the free field \eqref{Sfree}, and then the term explicitly written out is its positive frequency part. Staruszkiewicz demands that this part, as in the Minkowski space theory, annihilates the vacuum. This fixes the representation uniquely.

As mentioned before, the theory treats the long range degrees of freedom per se, and it does not seem possible to extend it to a consistent theory of quantum fields in the whole spacetime, with properly represented translations and the corresponding energy-momentum. However, what makes the theory intriguing is the fact that its structure shows critical dependence on the value of the fine structure constant $\alpha=e^2/(\hbar c)$ (we write here all constants explicitly to make it clear that the dimensionless constant is concerned). Namely, let $\W$ be the vacuum state vector, and consider the closed linear span $\Hc_e$ of all vectors of the form $\pi\big(W(0,c_v)\big)\W$, for all four-velocities $v$.
This subspace of the whole representation space is invariant under the Lorentz group, thus it carries its representation. Using the methods of Gelfand et al.\ \cite{gel66}, Staruszkiewicz succeeded in decomposing this representation into irreducibles, with the following remarkable result \cite{sta92}: for $z\equiv\alpha/\pi>1$ it decomposes into a continuous direct integral of main series irreducible representations $\mathfrak{S}_{0,\rho}$ (all of them with zero value of the second Casimir operator), while for $z<1$ there is one discrete addition of a representation $\mathfrak{D}_\nu$ from the supplementary series, with $\nu=1-\sqrt{z}$, which corresponds to the eigenvalue $z(2-z)$ of the first Casimir operator  (see also \cite{sta04}). The critical value $\alpha=\pi$ seems far from physical significance, but maybe its actual physical value is hidden in the theory in a more subtle way? This, as it seems, is the hope of the author of the theory.

\section{Conclusion}
There are various levels, at which infrared problem of quantum electrodynamics may be understood, and addressed. Within the limits of local theory, both the practical problem of dealing with the so called infrared infinities in Feynman diagrams, as well as that of constructing scattering operator, are discussed in another contribution to this Encyclopedia \cite{duc23}. Here, we described approaches which, at the cost of sacrificing unrestricted validity of locality, open new perspectives on the fundamental questions formulated at the beginning of the article in Abstract. Mathematically precise investigations in this direction have been pursued for  at least three decades, while recently more heuristic ideas, some of them based on rediscovered properties known before, are becoming popular. Here we have stressed the necessity of precise mathematical language for reliable conclusions. We have reported preliminary results of perturbation calculation in the quantum framework extended to nonlocal, asymptotic variables in spacelike infinity. The main problem for further developments will be the question of ultraviolet structure. In Staruszkiewicz's theory, the main problem is the question whether it contains further characteristic values of the fine structure constant.

\end{document}